\documentclass[preprint,preprintnumbers,amssymb,prd,nofootinbib,tightenlines]{revtex4}
\usepackage{graphicx}
\usepackage{bm}

\begin{document}
     
\baselineskip=15pt
\parskip=5pt
   

\hfill UK/TP-2003-07 
 
\hfill SMU-HEP-03-10 
 
\vspace*{0.7in}  

\title{Observing Direct CP Violation in Untagged $\bm{B}$-Meson Decays}  
   
\author{Susan Gardner${}^{1}$}   
\email{gardner@pa.uky.edu}
   
\author{Jusak Tandean${}^{1,2}$}  
\email{jtandean@mail.physics.smu.edu}   

\affiliation{
${}^1$Department of Physics and Astronomy, University of Kentucky, 
Lexington, Kentucky 40506-0055
\vspace{1ex} \\
}

\affiliation{  
${}^2$Department of Physics, \\ 
Southern Methodist University,   
Dallas, Texas 75275-0175\footnote{Present address.}  
\vspace{1ex} \\
}  


\begin{abstract}
Direct CP violation can exist in untagged $B$-meson decays to 
self-conjugate, three-particle final states; it would be realized
as a population asymmetry in the untagged decay rate 
across the mirror line 
of the Dalitz plot of the three-body decay.   
We explore the numerical size of this
direct CP-violating effect in a variety of $B$-meson 
decays to three pseudoscalar mesons; 
we show that the resulting asymmetry is comparable to
the partial rate asymmetry in the analogous tagged decays, making
the search for direct CP violation in the untagged decay rate, 
for which greater statistics accrue, advantageous. 
\end{abstract}


\maketitle

\section{Introduction}   

The recent observation of CP violation in the $B$-meson system, 
by dint of the measurement of a nonzero, 
CP-violating asymmetry in the decay 
$\,B^0(\bar B^0)\to J/\psi K_S\,$ and related modes~\cite{expt},
opens a new era of 
discovery. At issue is whether or not 
CP violation is realized 
exclusively through a single phase, $\delta_{KM}$, in 
the elements of the Cabibbo-Kobayashi-Maskawa 
(CKM) matrix~\cite{ckm}, as in the standard model (SM). 
The observed asymmetry, 
realized through
the interference of  $B^0$-$\bar B^0$ mixing and direct decay, 
is in accord with SM
expectations~\cite{ckmfit}, but 
additional, critical tests of 
the SM picture 
have as yet to be realized. For example, 
the $\,B^0(\bar B^0)\to J/\psi K_S\,$ result, coupled with the hierarchical
nature of the CKM matrix~\cite{wolfenstein}, suggests that significant 
direct CP-violating effects ought to exist in the $B$-meson system. 
Moreover, the determined angles of the unitarity triangle, whatever
they may be, must be universal, and they must sum to a multiple of
$\pi$~\cite{NQ}. In this paper, 
we study a new method of elucidating 
the presence of direct CP violation. The empirical observation of 
direct CP violation in the $B$ system would falsify models in which
CP violation is associated overwhelmingly with $\,|\Delta B|=2\,$, 
rather than $\,|\Delta B|=1\,$, processes: such models are 
``essentially'' superweak~\cite{superweak,superweak',superweak2,bigiplb}.
The observation of direct CP violation in the 
$B$-meson system is needed to clarify the nature and origin of CP
violation in nature. 
   
Direct CP violation (DCPV) in the $B$-meson system is 
realized if 
$|\bar A_{\bar f}/ A_{f}| \ne 1$, where $A_f$ is the decay
amplitude for $B\to f$ and $\bar A_{\bar f}$ is the decay amplitude
for the CP-conjugate process $\bar B\to \bar f$; it 
can be established in a variety of ways. 
Let us consider the possibilities.
Perhaps most familiar is the partial-rate asymmetry, ${\cal A}_{\rm CP}$, 
for which 
$\,{\cal A}_{\rm CP}^{}\propto\Gamma\bigl(B\to f \bigr)-
\Gamma\bigl(\bar B\to\bar f \bigr),\,$ 
with $\Gamma\bigl(B\to f \bigr) \propto |A_{f}|^2$. 
No conclusive evidence for DCPV has been found thus 
far, although early results 
in certain modes have been quite suggestive~\cite{dcpv_x1,dcpv_x2,dcpv_x3}. 
Another pathway to DCPV is realized through the
comparison of $a_{\rm CP}(f)$, the CP-violating asymmetry associated with the
interference of $B^0- \bar B^0$ mixing and direct decay, for two 
different self-conjugate final states $f$. Were 
$|a_{\rm CP}(f)| - a_{\rm CP}(\psi K_S)$ nonzero for a final state $f$, 
one would establish the existence of DCPV, as 
in the manner of the $\epsilon'$ parameter in the neutral-kaon system.  
Indeed, such a difference has been termed $\epsilon_{B}'$~\cite{eB'}. 
Experimentally, there is no conclusive evidence, as yet, for a nonzero
$\epsilon_{B}'$~\cite{a(f),a(f)more}.
It is worth noting 
that $\epsilon_{B}'$ can vanish 
for a specific final state, such 
as $\pi^+ \pi^-$, even if DCPV is 
present in nature~\cite{winstein,eB'plus}. 
Direct CP violation can also be established through the study of
the angular distribution of  $B$  decays into 
two vector mesons~\cite{valencia,Kayser:ww,Dunietz:1990cj,Kramer:1991xw,sinha2}, 
though no empirical limits on the DCPV terms exist as yet. 
Finally, DCPV can be established through a population asymmetry
in the Dalitz plot associated with the untagged decay rate
into self-conjugate final states~\cite{gardner,burdman}, as we will 
develop in detail. The last two methods have common elements;
in particular, 
neither method requires tagging nor time-dependence 
to find DCPV in the neutral $B$-meson system.

In this paper, we study the method proposed in 
Ref.~\cite{gardner} to search for DCPV 
in the decays of neutral, heavy mesons. 
The method uses untagged, neutral meson decays into 
self-conjugate final-states~\cite{Carter,Bigi,Dunietz:1986vi}
containing more than two hadrons\footnote{A note on language: 
a two-particle, ``self-conjugate'' state
is a CP-eigenstate, whereas a multi-particle ($n>2$) state is not 
a CP-eigenstate, in general. Nevertheless, we shall persist in 
using the phrase ``self-conjugate'' to describe multi-particle states
which are ``CP self-conjugate states
in particle content''~\cite{babarbook}.}.
The multi-particle ($n > 2$) final states realized in 
heavy-meson decays possess a rich resonance structure, 
entailing the possibility 
of detecting DCPV 
without tagging the flavor of the 
decaying neutral meson. 
For example, 
DCPV can occur if we can separate the self-conjugate final state, via
the resonances which appear, 
into distinct, CP-conjugate states. 
This condition finds it analogue in stereochemistry: we refer
to molecules which are non-superimposable, mirror images of each
other as enantiomers~\cite{ch41}, so that we term 
non-superimposable, 
CP-conjugate states as CP-enantiomers~\cite{gardner}. 
In $B\to \pi^+\pi^-\pi^0$ decay, e.g., 
the intermediate 
states $\rho^+\pi^-$ and $\rho^-\pi^+$ form CP-enantiomers, as
they are distinct, CP-conjugate states. As a result, 
the untagged decay rate contains a CP-odd amplitude combination.
The empirical presence of this CP-odd interference term in the untagged
decay rate would be realized in the Dalitz plot as a population asymmetry, 
reflective of direct CP violation~\cite{gardner}. 
The use of untagged decays to search for direct CP violation 
is also possible in $B\to V_1 V_2$ decays; 
its practical advantage is that 
there is no loss of statistics due to tagging.  

We focus in this paper on the application of this method to untagged 
$B$-meson decays into self-conjugate
final states of three pseudoscalar mesons. 
In Sec.~\ref{dcpv}, we detail, after Ref.~\cite{gardner}, 
how direct 
CP violation can arise in untagged, neutral-$B$-meson decays. 
In Sec.~\ref{cf} 
we discuss how the expected size of the population asymmetry 
compares, on general grounds, 
to that of the analogous partial-rate
asymmetry. 
In Sec.~\ref{results}, we make explicit numerical estimates
of the population asymmetries in particular channels and 
discuss strategies for their experimental realization.   
In specific, we consider the decays   
$\,B_d^{}\to\pi^+\pi^-\pi^0,\,D^+D^-\pi^0\,$, as well as 
the analogous decays 
$\,B_s^{}\to K^+K^-\pi^0,\,D_s^+D_s^-\pi^0\,$  in the $B_s$ system. 
These final states contain the 
CP-enantiomers $\rho^\pm\pi^\mp$,  $D^{*\pm}D^\mp$,  $K^{*\pm}K^\mp$, 
and $D_s^{*\pm}D_s^\mp$, respectively. We
conclude with a summary and accompanying outlook 
in Sec.~\ref{conclusion}.  

\section{Direct CP violation in untagged $\bm{B}$ decays\label{dcpv}}    
  
\subsection{\mbox{\boldmath$B\to \rho\pi \to \pi^+\pi^- \pi^0$}}
\label{dcpva}

We follow Ref.~\cite{gardner} and use the decay  
$\,B\to\rho\pi\to\pi^+\pi^-\pi^0\,$ as a paradigm 
to illustrate 
how direct  CP violation can occur in untagged, neutral heavy-meson 
decays.   
We express the amplitudes for  
$\,B^0 \,(\bar B^0)\,\to\rho^+\pi^-,\rho^-\pi^+,\rho^0\pi^0\,$ decay as   
\begin{eqnarray}   \label{M_pirho} 
\begin{array}{c}   \displaystyle   
{\cal M}_{\rho^k\pi^l}^{}  \,=\,  
a_{kl}^{}\, \varepsilon_\rho^*\cdot p_\pi^{}   \,\,     
\hspace{1em}  \hbox{and} \hspace{1em}  
\bar {\cal M}_{\rho^k\pi^l}^{}  \,=\,     
 \bar a_{kl}^{}\, \varepsilon_\rho^*\cdot p_\pi^{}   \,\,,  
\end{array}     
\end{eqnarray}     
respectively. 
Under an assumption of $\rho$ dominance, the amplitudes for  
$\,B^0\, (\bar B^0) \to\pi^+\bigl(p_+^{}\bigr)\, 
\pi^-\bigl(p_-^{}\bigr)\, \pi^0\bigl(p_0^{}\bigr)\,$  decay  
can be written as~\cite{SnyQui,QuiSil}
\begin{eqnarray}   \label{M_3pi} 
\begin{array}{c}   \displaystyle   
{\cal M}_{3\pi}^{}  \,=\,   
a_{+-}^{}\, f_+^{} + a_{-+}^{}\, f_-^{} + a_{00}^{}\, f_0^{}   
\,=\,  a_g^{}\, f_g^{} + a_u^{}\, f_u^{} + a_n^{}\, f_n^{}   \,\,,      
\vspace{2ex} \\   \displaystyle   
\bar {\cal M}_{3\pi}^{}  \,=\,  
 \bar  a_{+-}^{}\, f_+^{}  + \bar a_{-+}^{}\, f_-^{} 
+  \bar a_{00}^{}\, f_0^{}   
\,=\,  
 \bar  a_g^{}\, f_g^{} + \bar a_u^{}\, f_u^{} + \bar a_n^{}\, f_n^{}   \,\,,  
\end{array}     
\end{eqnarray}     
where  we have summed over the $\rho$ polarization. We define 
\begin{eqnarray}   \label{aba}
\begin{array}{c}   \displaystyle   
a_g^{}  \,=\,  a_{+-}^{}+a_{-+}^{}   \,\,,   
\hspace{3em}  
a_u^{}  \,=\,  a_{+-}^{}-a_{-+}^{}   \,\,,   
\hspace{3em}  
a_n^{}  \,=\,  2\, a_{00}^{}   \,\,,      
\vspace{1ex} \\   \displaystyle   
\bar a_g^{}  \,=\,  \bar a_{+-}^{}+\bar a_{-+}^{}   \,\,,   
\hspace{3em}  
\bar a_u^{}  \,=\,  \bar a_{+-}^{}-\bar a_{-+}^{}   \,\,,   
\hspace{3em}  
\bar a_n^{}  \,=\,  2\, \bar a_{00}^{}   \,\,,
\end{array}     
\end{eqnarray}     
and 
\begin{eqnarray}   \label{f_cdn}
\begin{array}{c}   \displaystyle   
f_{\pm}^{}  \,=\,   
\pm\mbox{$\frac{1}{2}$} \bigl( s_{+-}^{}-s_{\mp0}^{} \bigr) \, 
\Gamma_{\rho\pi\pi}^{} \bigl( s_{\pm0}^{} \bigr)   \,\,,   
\hspace{2em}  
f_0^{}  \,=\,   
\mbox{$\frac{1}{2}$} \bigl( s_{-0}^{}-s_{+0}^{} \bigr) \,   
\Gamma_{\rho\pi\pi}^{} \bigl( s_{+-}^{} \bigr)   \,\,,      
\vspace{2ex} \\   \displaystyle   
2\, f_g^{}  \,=\,  f_+^{}+f_-^{}   \,\,,   
\hspace{3em}  
2\, f_u^{}  \,=\,  f_+^{}-f_-^{}   \,\,,   
\hspace{3em}  
2\, f_n^{}  \,=\,  f_0^{}   \,\,, 
\end{array}     
\end{eqnarray}     
where we have neglected the charged-neutral pion mass difference. 
Note that $\Gamma_{\rho\pi\pi}^{}(s)$ is the $\rho\to\pi\pi$ 
vertex function, which can be determined from 
$e^+e^-\to\pi^+\pi^-$ data,
and that 
$\,s_{kl}^{}\equiv \bigl( p_k^{}+p_l^{} \bigr)^2 .\,$  
Let us consider the untagged decay width: 
\begin{eqnarray}   \label{Gamma_3pi}
\Gamma_{3\pi}^{}  \,=\,  
\int {{\rm d}s_{+0}^{}\,{\rm d}s_{-0}^{}\over 256\pi^3\, m_{B^0}^3}   
\left( \left| \vphantom{\bar M}{\cal M}_{3\pi}^{} \right| ^2 
      + \left| \bar{\cal M}_{3\pi}^{} \right| ^2 \right) 
\,=\,   
\Gamma_{3\pi}^{(1)} + \Gamma_{3\pi}^{(2)} + \Gamma_{3\pi}^{(3)}    \,\,.   
\end{eqnarray}     
We can separate this width into pieces which contain 
CP-even, as well as CP-odd,  amplitude combinations, so that 
\begin{eqnarray}  \label{Gamma1}   
\Gamma_{3\pi}^{(1)}  \,=\,   
\int {\rm d}\Phi \left\{ \sum_{\kappa=g,u,n} 
\left( \bigl| a_\kappa^{} \bigr|^2 + \bigl| \bar a_\kappa^{} \bigr|^2 
\vphantom{|_|^|} \right) \bigl| f_\kappa^{} \bigr|^2 
\,+\,  
2\, {\rm Re} \left[ 
\left( a_g^* a_n^{}+\bar a_g^* \bar a_n^{} \right) f_g^* f_n^{} 
\vphantom{|_|^|} \right]   
\right\} \,\,,   
\end{eqnarray}     
whereas
\begin{eqnarray}  \label{Gamma23}   
\Gamma_{3\pi}^{(2)}  \,=\,   
\int {\rm d}\Phi\; 2\, {\rm Re} \left[ 
\left( a_g^* a_u^{}+\bar a_g^* \bar a_u^{} \right) f_g^* f_u^{} 
\vphantom{|_|^|} \right]   \,\,,  
\hspace{2em}   
\Gamma_{3\pi}^{(3)}  \,=\,   
\int {\rm d}\Phi\; 2\, {\rm Re} \left[ 
\left( a_u^* a_n^{}+\bar a_u^* \bar a_n^{} \right) f_u^* f_n^{} 
\vphantom{|_|^|} \right]   \,\,,   
\end{eqnarray}     
with 
\begin{eqnarray}  \label{dPS}   
{\rm d}\Phi  \,\equiv\,  
{{\rm d}s_{+0}^{}\,{\rm d}s_{-0}^{}\over 256\pi^3\, m_{B^0}^3}   \,\,.  
\end{eqnarray}     
It is worth emphasizing that the $f_k$ are known functions, as
$\Gamma_{\rho\pi\pi}^{}(s)$ itself can be determined from 
data. Consequently, the coefficients of $f_k^* f_l^{}$ are observables: 
they can, in principle,
be determined from the Dalitz plot of the 
$\pi^+\pi^-\pi^0$  final state~\cite{SnyQui,QuiSil}.

\begin{figure}[ht]
\includegraphics[width=3in]{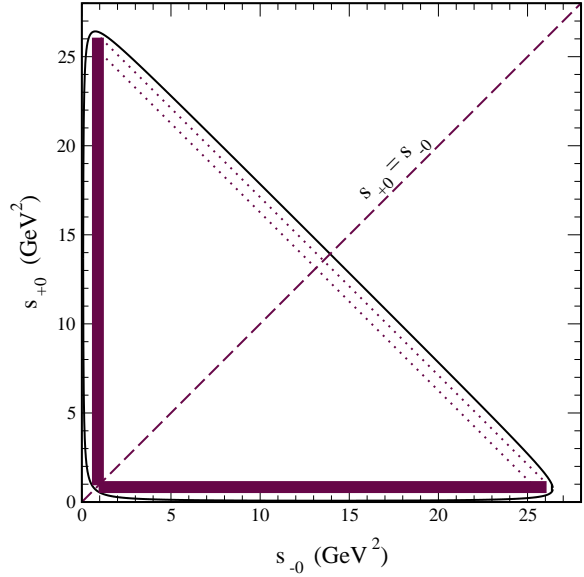}   
\vspace{-0.2cm}
\caption{\label{Bto3pi} 
The allowed Dalitz region for 
$\,B^0,\bar B^0\to\pi^+\pi^-\pi^0\,$ decay.
The decays mediated by the $\rho^\pm \pi^\mp$ intermediate states 
dominate the Dalitz plot; their location is indicated schematically
by the shaded bands. The $\rho^0$ band, reflective of the 
$\rho^0 \pi^0$ intermediate state, is indicated schematically 
by the dotted lines. An asymmetry in the population of the Dalitz plot 
about the $s_{+0}=s_{-0}$ ``mirror line'' signals 
direct CP violation. Note that 
the asymmetries ${\cal A}_{3\pi}^{(2)}$ and ${\cal A}_{3\pi}^{(3)}$, as
given in Eq.~(\protect{\ref{asymobs}}), 
describe the population asymmetry of the $\rho^\pm$ bands 
and that of the overlapping $\rho^\pm$-$\rho^0$ bands
about the mirror line, respectively. 
}   
\end{figure}

We now consider the transformation properties of 
the $a_{ij}$ amplitudes under CP. 
Under the conventions we detail in Sec.~\ref{dcpvb}, we have 
\begin{eqnarray}   
\label{convdef}
a_{+-}^{}  \,\stackrel{\,CP}{\longleftrightarrow}\,  
\bar a_{-+}^{}   \,\,,  
\hspace{2em}  
a_{-+}^{}  \,\stackrel{\,CP}{\longleftrightarrow}\,  
\bar a_{+-}^{}   \,\,,  
\hspace{2em}  
a_{00}^{}  \,\stackrel{\,CP}{\longleftrightarrow}\,  
\bar a_{00}^{}   \,\,,     
\end{eqnarray}     
which translate into   
\begin{eqnarray}   \label{CP[a_cdn]}
a_g^{}  \,\stackrel{\,CP}{\longleftrightarrow}\,  +\bar a_g^{}   \,\,,  
\hspace{2em}  
a_u^{}  \,\stackrel{\,CP}{\longleftrightarrow}\,  -\bar a_u^{}   \,\,,  
\hspace{2em}  
a_n^{}  \,\stackrel{\,CP}{\longleftrightarrow}\,  +\bar a_n^{}   \,\,.     
\end{eqnarray}     
It follows that 
the combination  
$\,a_g^* a_n^{}+\bar a_g^* \bar a_n^{}\,$  in Eq.~(\ref{Gamma1}) 
is CP even,  whereas  the combinations 
\begin{eqnarray}  \label{ac*ad}   
a_g^* a_u^{}+\bar a_g^* \bar a_u^{}  
\hspace{2em} {\rm and} \hspace{2em}   
a_u^* a_n^{}+\bar a_u^* \bar a_n^{}   
\end{eqnarray}     
in Eq.~(\ref{Gamma23}) are CP odd, as first noted by Quinn
and Silva~\cite{QuiSil}. The discernable presence of either
of the last two combinations signals direct CP violation. 
These quantities can be measured with the aid of 
the  Dalitz plot of  $s_{+0}^{}$  vs. $s_{-0}^{}$, as illustrated in 
Fig.~\ref{Bto3pi}. To see how this is useful, note that 
under $p_+ \leftrightarrow p_-$ we have, from Eq.~(\ref{f_cdn}), that  
\begin{eqnarray}  \label{fpm}   
f_{g,n}^{} \bigl(s_{+0}^{},s_{-0}^{}\bigr)  \,=\,  
-f_{g,n}^{} \bigl(s_{-0}^{},s_{+0}^{}\bigr)   \,\,,  
\hspace{2em} 
f_u^{} \bigl(s_{+0}^{},s_{-0}^{}\bigr)  \,=\,  
+f_u^{} \bigl(s_{-0}^{},s_{+0}^{}\bigr)   \,\,,  
\end{eqnarray}     
which implies that  $f_g^*f_u^{}$  and  $f_u^*f_n^{}$  
are both odd under the interchange   
$\,s_{+0}^{}\leftrightarrow s_{-0}^{},\,$  
whereas  $f_g^*f_n^{}$  is even. Thus the CP-odd terms make
no contribution whatsoever to the total, untagged decay rate; the
behavior of Eq.~(\ref{fpm}) ensures that these contributions 
vanish identically in the integral over phase space. However, 
they do generate a population asymmetry 
about the  $\,s_{+0}^{}=s_{-0}^{}\,$  line in the $3\pi$ Dalitz plot. 
Explicitly, noting Eqs.~(\ref{Gamma1})  and (\ref{Gamma23}), we have 
\begin{eqnarray}  \label{Gammai}   
\Gamma_{3\pi}^{(1)} \bigl[s_{+0}^{}\mbox{$>$}s_{-0}^{}\bigr] 
\,=\,   
+\Gamma_{3\pi}^{(1)} \bigl[s_{+0}^{}\mbox{$<$}s_{-0}^{}\bigr]   \,\,,  
\hspace{2em}  
\Gamma_{3\pi}^{(2,3)} \bigl[s_{+0}^{}\mbox{$>$}s_{-0}^{}\bigr] 
\,=\,   
-\Gamma_{3\pi}^{(2,3)} \bigl[s_{+0}^{}\mbox{$<$}s_{-0}^{}\bigr]   \,\,,  
\end{eqnarray}     
where the inequalities within the square 
brackets indicate the regions of integration within the 
Dalitz plot.    
The corresponding asymmetry can be expressed as  
\begin{eqnarray}  \label{a_3pi}   
{\cal A}_{3\pi}^{}  \,\equiv\,  
{ \Gamma_{3\pi}^{} \bigl[s_{+0}^{}\mbox{$>$}s_{-0}^{}\bigr] 
 - \Gamma_{3\pi}^{} \bigl[s_{+0}^{}\mbox{$<$}s_{-0}^{}\bigr]  
 \over 
 \Gamma_{3\pi}^{} \bigl[s_{+0}^{}\mbox{$>$}s_{-0}^{}\bigr] 
 + \Gamma_{3\pi}^{} \bigl[s_{+0}^{}\mbox{$<$}s_{-0}^{}\bigr] }  
\,=\,   
{ \Gamma_{3\pi}^{(2)} \bigl[s_{+0}^{}\mbox{$>$}s_{-0}^{}\bigr] 
 + \Gamma_{3\pi}^{(3)} \bigl[s_{+0}^{}\mbox{$>$}s_{-0}^{}\bigr]  
 \over 
 \Gamma_{3\pi}^{(1)} \bigl[s_{+0}^{}\mbox{$>$}s_{-0}^{}\bigr] }   \,\,.  
\end{eqnarray}     
A nonvanishing value of  ${\cal A}_{3\pi}^{}$  reflects 
direct CP violation in the  $\,B\to\rho\pi\,$  amplitudes. 
The individual asymmetries, 
\begin{equation}
\label{asymobs}
{\cal A}_{3\pi}^{(2)}\equiv 
\frac{ \Gamma_{3\pi}^{(2)} \bigl[s_{+0}^{}\mbox{$>$}s_{-0}^{}\bigr] }{
 \Gamma_{3\pi}^{(1)} \bigl[s_{+0}^{}\mbox{$>$}s_{-0}^{}\bigr] }   
\,\,,  \hspace{3em}  
{\cal A}_{3\pi}^{(3)}\equiv 
 \frac{\Gamma_{3\pi}^{(3)} \bigl[s_{+0}^{}\mbox{$>$}s_{-0}^{}\bigr]}  
{\Gamma_{3\pi}^{(1)} \bigl[s_{+0}^{}\mbox{$>$}s_{-0}^{}\bigr] }   \,\,,  
\end{equation}
are distinguishable through their location in 
the Dalitz plot; moreover, they are proportional 
to different combinations of weak amplitudes, as clear from 
Eq.~(\ref{Gamma23}). The $f_k$ functions embedded in the definitions 
of $\Gamma_{3\pi}^{(2,3)}$ ensure that the population asymmetry is 
restricted to the $\rho^k$ bands. In specific, 
$\Gamma_{3\pi}^{(2)}$ is nonzero if the population of the
$\rho^+$ band for $s_{-0} > s_{+0}$ is different from that of the
$\rho^-$ band for $s_{-0} < s_{+0}$, whereas $\Gamma_{3\pi}^{(3)}$ 
is nonzero if the population of the $\rho^+$-$\rho^0$ interference 
region, which occurs for $s_{-0} > s_{+0}$, is different from that of the 
$\rho^-$-$\rho^0$ interference region, which occurs for 
$s_{-0} < s_{+0}$. 
It turns out that 
$\Gamma_{3\pi}^{(3)}$ is sufficiently
small, on general grounds, that its impact on $\Gamma_{3\pi}^{(2)}$ 
is negligible. 

\vspace{0.5cm}

We have illustrated how DCPV can occur in the untagged decay rate. 
The untagged decay rate 
is distinct from the untagged combination
of time-dependent rates. To appreciate the consequences of this, 
we consider the time dependence of neutral, $B$-meson decay. 
We let $|{B^0(t)}\rangle$ denote the 
state which is tagged as a $B^0$ meson at time $t=0$. 
We define the mass eigenstates     
under the weak interaction as
$\,|B_H^{}\rangle =p | B^0 \rangle -q | \bar B^0\rangle \,$  
and  $\,|B_L^{}\rangle =p | B^0 \rangle + q | \bar B^0\rangle \,$; we 
denote their masses and widths by $M_{H,L}^{}$ and    
$\Gamma_{H,L}^{}$, respectively. 
Denoting ${\cal M}_f(t)\,\, \bigl(\bar {\cal M}_f(t)\bigr)$ as the decay 
amplitude for $\,B^0(t)\,\, \bigl(\bar B^0(t)\bigr) \to f\,$  decay,  
and assuming mixing CP-violation to be  
negligible, so that $\,|q/p|=1\,$, we have~\cite{bigirev,Anikeev:2001rk}
\begin{eqnarray}   \label{B0(t)->f}      
\left| {\cal M}_f(t) \vphantom{\bar M} \right| ^2  &=&  
{\rm e}^{-\Gamma t} \left[  
\mbox{$\frac{1}{2}$}  
\left( \bigl| {\cal M}_f^{} \bigr|^2+\bigl| \bar{\cal M}_f^{} \bigr|^2 
\vphantom{|_|^|} \right) \cosh {\Delta\Gamma\, t\over 2} 
\,+\,    
\mbox{$\frac{1}{2}$}  
\left( \bigl| {\cal M}_f^{} \bigr|^2-\bigl| \bar{\cal M}_f^{} \bigr|^2 
\vphantom{|_|^|} \right) \cos(\Delta M\, t)   
\right.  
\nonumber \\ && \left. \hspace*{0.4in}     
+\; 
{\rm Re} \left( {q\over p} {\cal M}_f^* \bar{\cal M}_f^{} \right) \, 
\sinh{\Delta\Gamma\, t\over 2}  
- {\rm Im} \left( {q\over p} {\cal M}_f^* \bar{\cal M}_f^{} \right) 
\sin(\Delta M\, t)   
\right]   \,\,,   
\end{eqnarray}   
and
\begin{eqnarray}   \label{bB0(t)->f}
\left| \bar{\cal M}_f(t) \right| ^2  &=&  
{\rm e}^{-\Gamma t} \left[  
\mbox{$\frac{1}{2}$}  
\left( \bigl| {\cal M}_f^{} \bigr|^2+\bigl| \bar{\cal M}_f^{} \bigr|^2 
\vphantom{|_|^|} \right) \cosh{\Delta\Gamma\, t\over 2}  
\,-\,  
\mbox{$\frac{1}{2}$}  
\left( \bigl| {\cal M}_f^{} \bigr|^2-\bigl| \bar{\cal M}_f^{} \bigr|^2 
\vphantom{|_|^|} \right) \cos(\Delta M\, t)
\right.   
\nonumber \\ && \left. \hspace*{0.4in}    
+\; 
{\rm Re} \left( {q\over p} {\cal M}_f^* \bar{\cal M}_f^{} \right) 
\sinh{\Delta\Gamma\, t\over 2}  
\,+\,    
{\rm Im} \left( {q\over p} {\cal M}_f^* \bar{\cal M}_f^{} \right) 
\sin(\Delta M\, t)   
\right]   \;.
\end{eqnarray}   
We have used the conventions of Ref.~\cite{Anikeev:2001rk}, so that 
 $\,2\,\Gamma\equiv\Gamma_H^{}+\Gamma_L^{}\,$  
and $\,\Delta M\equiv M_H^{}-M_L^{},\,$  though we define
$\,\Delta\Gamma\equiv\Gamma_H^{}-\Gamma_L^{}.\,$  
The untagged combination of time-dependent rates is given by  
\begin{eqnarray}   \label{B0(t)+bB0(t)->f}      
\left| {\cal M}_f(t) \vphantom{\bar M} \right| ^2  
+ \left| \bar{\cal M}_f(t) \right| ^2  
\,=\,  
{\rm e}^{-\Gamma t} \left[  
\left( \bigl| {\cal M}_f^{} \bigr|^2+\bigl| \bar{\cal M}_f^{} \bigr|^2 
\vphantom{|_|^|} \right) \cosh {\Delta\Gamma\, t\over 2} 
+ 2\, {\rm Re} \left( {q\over p}{\cal M}_f^* \bar{\cal M}_f^{} \right) \, 
\sinh{\Delta\Gamma\, t\over 2}   
\right]   \;.  
\end{eqnarray}   
The population asymmetry associated with the untagged combination
of time-dependent rates need not be a signal of 
direct CP violation at {\it nonzero}
$t$. The population asymmetries 
are generated by terms of form 
$f_g^* f_u^{} $ or $f_u^* f_n^{}$, as in Eqs.~(\ref{Gamma23},\ref{asymobs});
at  $\,t\neq0,\,$  ${\cal M}_f^* \bar{\cal M}_f^{}$ enters as well. 
Such terms are associated with 
CP-odd amplitudes in ${\rm Re}({\cal M}_f^* \bar{\cal M}_f^{})$, 
but with CP-even amplitudes in ${\rm Im}({\cal M}_f^* \bar{\cal M}_f^{})$. 
The pieces of interest in ${\cal M}_f^* \bar{\cal M}_f^{}$ are of form 
\begin{equation}
{\cal M}_f^* \bar{\cal M}_f^{}\Big|_{{}_{f_g^* f_u^{}}}
\,=\,  a_{gu}^{+}\, \mbox{Re}(f_g^\ast f_u^{}) + 
{\rm i}\, a_{gu}^{-}\, \mbox{Im}(f_g^\ast f_u^{}) 
\end{equation}
where $a_{gu}^{\pm}= a_g^\ast \bar a_u^{} \pm a_u^\ast \bar a_g^{}$ and the
replacement $g,u \leftrightarrow u,n$ yields the $f_u^* f_n^{}$ terms. 
We note 
\begin{equation}
a_{gu}^{\pm} \,\stackrel{\,CP}{\longleftrightarrow}\,  
\mp a_{gu}^{\pm\,*}\,\,,
\end{equation}
so that the
CP properties of ${\rm (Re,\,Im)}({\cal M}_f^* \bar{\cal M}_f^{})$ follow. 
Since $q/p$ is complex, 
${\rm Im}({\cal M}_f^* \bar{\cal M}_f^{})$ enters 
${\rm Re} ( q{\cal M}_f^* \bar{\cal M}_f^{}/p)$ as well, so that the population
asymmetry associated with this $t$-dependent term can be generated 
by direct CP violation, but does not occur exclusively because of it. 
In a similar vein, we conclude that a population asymmetry in 
$| {\cal M}_f(t) \vphantom{\bar M}| ^2  
- | \bar{\cal M}_f(t) | ^2$ need not be reflective of direct CP violation; 
at $t=0$ it is decidedly not. 
Since $\Delta \Gamma \ll \Gamma$ in the SM~\cite{Bigi:1986vr}, though
it is likely more significant in 
the $B_s$ system~\cite{Aleksan:1993qp,Beneke:1996gn}\footnote{We note 
$\Delta \Gamma_d^{\rm SM}/ \Gamma_d \approx 3 \times 10^{-3}$, 
whereas $\Delta \Gamma_s^{\rm SM}/ \Gamma_s =0.12 \pm 0.06$, as per
Ref.~\cite{Anikeev:2001rk} and references therein.}, 
the practical impact of this observation ought to be minor; it 
nevertheless remains that the population asymmetry signals 
direct CP violation in the untagged decay rate, rather than in the
untagged combination of tagged, time-dependent rates. 
However, if the latter quantity is integrated in $t$ over a symmetric
region about $t=0$, the undesired term vanishes, and the resulting
time-integrated, population asymmetry is also reflective of direct 
CP violation.

\subsection{\mbox{\boldmath{$B\to \pi^+\pi^- \pi^0$}}}
\label{dcpvb}

Herewith we demonstrate that the population asymmetry in the untagged
combination of time-independent rates in $B\to \pi^+ \pi^-\pi^0$ 
decay is a signature of direct CP violation; we need not confine
ourselves to the regions of the Dalitz plot dominated by the $\rho$
resonance. Indeed, our demonstration follows for all three-body decays 
to final states which are self-conjugate in particle content 
and whose particles are spinless;
the final-state particles need not be related by isospin symmetry. 

Let us consider the amplitude for 
$B\to \pi^+ \pi^-\pi^0$ decay. We enumerate the possible 
contributions by the orbital angular momentum of the final-state 
mesons, as this suffices to characterize the CP of the
final, three-particle state. 
We note, in passing, that 
a complete parametrization of the decay of a pseudoscalar meson 
to $\pi^+\pi^-\pi^0$ in terms of the 
isospin $I$ of the pions and $|\Delta I|$ of the weak transition
can be found in Ref.~\cite{zemach}. 
The final state must have zero total angular momentum, so 
that we write  $\bigl(\pi_1^{}\pi_2^{}\bigr)_{l} \bigl(\pi_3^{}\bigr)_{l}$,  
where $l$  denotes the total orbital angular momentum quantum number of  
the particles in parentheses, combined to yield a state of 
total angular momentum zero.\footnote{Note that  $(\pi_3^{})_{l}^{}$  
denotes the orbital angular momentum of $\pi_3^{}$ with respect to 
the center-of-mass of the  $\pi_1^{}\pi_2^{}$  system.}  
We have implicitly summed over the magnetic
quantum numbers, $m$, possible for fixed $l$. 
The decay mechanism can distinguish the way in which the three 
pions are coupled to yield a state of total angular momentum zero. 
For example, in decays mediated by a two-body intermediate state,
such as 
$\,B\to\rho^+\pi^-\to \bigl(\pi^+\pi^0\bigr)_{1} \bigl(\pi^-\bigr)_{1}\,$ 
or 
$\,B\to\rho^-\pi^+\to \bigl(\pi^- \pi^0\bigr)_{1} \bigl(\pi^+\bigr)_{1},\,$ 
the differing ways in which two of the three pions are coupled to a  $\,l=1\,$  
state reflect distinct weak amplitudes. 
We denote the amplitude for 
$\,B\to\bigl(\pi^+ \pi^0\bigr)_{1} \bigl(\pi^-\bigr)_{1}$ decay 
as  ${\cal M}_{+-}^{}\bigl(p_+^{},p_-^{},p_0^{}\bigr)$, so that the
first subscript denotes the net charge of two pions coupled to a $l=1$
state and the second subscript denotes the charge of the third pion;  
its arguments $p_+,p_-,p_0$ denote the momenta of the 
$\pi^+,\pi^-$, and $\pi^0$ mesons, respectively, so that we 
define 
\begin{eqnarray}   \label{M_+-^l} 
\begin{array}{c}   \displaystyle   
{\cal M}_{\pm\mp}^{l}\bigl(p_+^{},p_-^{},p_0^{}\bigr)  \,=\,   
\Bigl\langle \bigl[\pi^\pm(p_\pm) \pi^0(p_0)\bigr]_l\, \pi^\mp(p_\mp)_l 
\Bigr| 
{\cal H}_{\rm eff}^{} \Bigl| B^0 \Bigr\rangle 
\,=\,  
\sum_i^{} \lambda_i^{*}\, \Bigl\langle \bigl(\pi^\pm\pi^0\bigr)_l\, 
\pi_l^\mp \Bigr| Q_i^\dagger \Bigl| B^0 \Bigr\rangle    \,\,,  
\vspace{2ex} \\   \displaystyle   
{\cal M}_{00}^{l}\bigl(p_+^{},p_-^{},p_0^{}\bigr)  \,=\,   
\Bigl\langle \bigl[\pi^+(p_+)\pi^-(p_-)\bigr]_l\, \pi^0(p_0)_l \Bigr| 
{\cal H}_{\rm eff}^{} \Bigl| B^0 \Bigr\rangle 
\,=\,  
\sum_i^{} \lambda_i^{*}\, \Bigl\langle \bigl(\pi^+\pi^-\bigr)_l\, 
\pi_l^0 \Bigr| Q_i^\dagger \Bigl| B^0 \Bigr\rangle    \,\,,  
\end{array}     
\end{eqnarray}     
with analogous quantities for $\bar B_0$ decay 
\begin{eqnarray}   \label{M_00^l} 
\begin{array}{c}   \displaystyle   
\bar{\cal M}_{\mp\pm}^{l}\bigl(p_+^{},p_-^{},p_0^{}\bigr)  \,=\,   
\Bigl\langle \bigl[\pi^\mp(p_\mp)\pi^0(p_0)\bigr]_l\, \pi^\pm(p_\pm)_l \Bigr| 
{\cal H}_{\rm eff}^{} \Bigl| \bar{B}^0 \Bigr\rangle  
\,=\,  
\sum_i^{} \lambda_i^{}\, \Bigl\langle \bigl(\pi^\mp\pi^0\bigr)_l\, 
\pi_l^\pm \Bigr| Q_i^{} \Bigl| \bar{B}^0 \Bigr\rangle    \,\,, 
\vspace{2ex} \\   \displaystyle   
\bar{\cal M}_{00}^{l}\bigl(p_+^{},p_-^{},p_0^{}\bigr)  \,=\,   
\Bigl\langle \bigl[\pi^+(p_+)\pi^-(p_-)\bigr]_l\, \pi^0(p_0)_l \Bigr| 
{\cal H}_{\rm eff}^{} \Bigl| \bar{B}^0 \Bigr\rangle  
\,=\,  
\sum_i^{} \lambda_i^{}\, \Bigl\langle \bigl(\pi^+\pi^-\bigr)_l\, 
\pi_l^0 \Bigr| Q_i^{} \Bigl| \bar{B}^0 \Bigr\rangle    \,\,, 
\end{array}     
\end{eqnarray}     
where we write the $|\Delta B|=1$ effective, 
weak Hamiltonian as  
\begin{eqnarray}   
{\cal H}_{\rm eff}^{}  \,=\,  
\sum_i^{} \Bigl( \lambda_i^{} Q_i^{}+\lambda_i^{*} Q_i^\dagger \Bigr)   \;.  
\end{eqnarray}     
To determine the transformation properties of these amplitudes
under CP, we have 
\begin{eqnarray}   \label{cp}   
\begin{array}{c}   \displaystyle   
{C}{P} |B\rangle  \,=\,  
\eta_B^{} \bigl| \bar B \bigr\rangle   \,\,,  
\hspace{3em}  
{C}{P} \bigl|\pi^0\bigr\rangle  \,=\,  
- \bigl| \pi^0 \bigr\rangle   \,\,,  
\hspace{3em}  
{C}{P} \bigl|\pi^+\bigr\rangle  \,=\,  
\eta_\pi^{} \bigl| \pi^- \bigr\rangle   \,\,,  
\vspace{2ex} \\   \displaystyle   
{C}{P}\, Q_i^{}\, 
\bigl({C}{P}\bigr)^\dagger 
\,=\,  \eta_Q^{}\, Q_i^\dagger   \,\,,    
\end{array}      
\end{eqnarray}      
where  $\eta_{B,\pi,Q}^{}$  are arbitrary phase-factors. Armed with
these conventions, we turn to the determination of the 
CP-properties of the three-meson
final state. Working in the rest frame of the two pions coupled
to angular momentum $l$, we note that 
\begin{equation}
{P} \Bigl| \bigl[ \pi_1^{}(\bm{p})\, \pi_2^{}(-\bm{p}) \bigr]_l\, 
\pi_3^{}(\bm{p}')_l^{} \Bigr\rangle
\,=\,
-\Bigl| \bigl[ \pi_1^{}(\bm{p})\, \pi_2^{}(-\bm{p}) \bigr]_l\, 
\pi_3^{}(\bm{p}')_l^{} \Bigr\rangle   \,\,,  
\end{equation}
where the minus sign emerges from the negative intrinsic parity of the 
pseudoscalar mesons. Under charge conjugation, 
\begin{equation}
{C} \Bigl| \bigl[ \pi^+(\bm{p})\, \pi^-(-\bm{p}) \bigr]_l\, 
\pi^0(\bm{p}')_l^{} \Bigr\rangle
\,=\,
\Bigl| \bigl[ \pi^-(\bm{p})\, \pi^+(-\bm{p}) \bigr]_l\, 
\pi^0(\bm{p}')_l^{} \Bigr\rangle
\,=\,
(-1)^l\, \Bigl| \bigl[ \pi^-(-\bm{p})\, \pi^+(\bm{p}) \bigr]_l\, 
\pi^0(\bm{p}')_l^{} \Bigr\rangle   \;.  
\end{equation}
It follows that  
\begin{eqnarray}   
{C}{P}   
\Bigl| \bigl[ \pi^+(\bm{p})\, \pi^-(-\bm{p}) \bigr]_l 
\pi^0(\bm{p}')_l^{} \Bigr\rangle
&\!=&\!
- {C} \Bigl| \bigl[ \pi^+(\bm{p})\, \pi^-(-\bm{p}) \bigr]_l 
\pi^0(\bm{p}')_l^{} \Bigr\rangle   
\,=\,
(-1)^{l+1}\, \Bigl| \bigl[ \pi^-(-\bm{p})\, \pi^+(\bm{p}) \bigr]_l 
\pi^0(\bm{p})_l^{} \Bigr\rangle   \,\,,   
\nonumber \\ 
\label{cp(3pi)}  
&& \hspace*{-5em}  
{C}{P} 
\Bigl| \bigl[ \pi^-(\bm{p})\,\pi^0(-\bm{p}) \bigr]_l\, 
\pi^+(\bm{p}')_l^{} \Bigr\rangle
\,=\,  
- \Bigl| \bigl[ \pi^+(\bm{p})\, \pi^0(-\bm{p}) \bigr]_l\,  
\pi^-(\bm{p}')_l^{} \Bigr\rangle   \,\,. 
\end{eqnarray}
We thus determine
\begin{eqnarray}   
\Bigl\langle \bigl[ \pi^-\bigl(p_-^{}\bigr)\, 
\pi^0\bigl(p_0^{}\bigr) \bigr]_l\, 
\pi^+\bigl(p_+^{}\bigr)_l \Bigr| Q_i^{} \Bigl| \bar{B}^0 \Bigr\rangle    
&=&  
\Bigl\langle 
\bigl[ \pi^-\bigl(p_-^{}\bigr)\, \pi^0\bigl(p_0^{}\bigr) \bigr]_l\, 
\pi^+\bigl(p_+^{}\bigr)_l \Bigr|  
\bigl({C}{P}\bigr)^\dagger\, 
{C}{P}\, Q_i^{}\, 
\bigl({C}{P}\bigr)^\dagger 
{C}{P} \Bigl| \bar{B}^0 \Bigr\rangle    
\nonumber \\
&=&  
-\eta_B^* \eta_Q^{}\, \Bigl\langle \bigl[ \pi^+\bigl(p_-^{}\bigr)\, 
\pi^0\bigl(p_0^{}\bigr) \bigr]_l\, \pi^-\bigl(p_+^{}\bigr)_l \Bigr| 
Q_i^\dagger\, \Bigl| B^0 \Bigr\rangle   \,\,,  
\end{eqnarray}     
where we work in the frame in which 
$\,\bm{p}_-^{}=\bm{p},\,$  $\,\bm{p}_0^{}=-\bm{p},\,$  and  
$\,\bm{p}_+^{}=\bm{p}'.\,$     
Moreover, 
\begin{eqnarray}   
\Bigl\langle \bigl(\pi^+\pi^-\bigr)_l\, 
\pi_l^0 \Bigr| Q_i^{} \Bigl| \bar{B}^0 \Bigr\rangle    
&=&  
\Bigl\langle \bigl(\pi^+\pi^-\bigr)_l\, \pi_l^0 \Bigr| 
\bigl({C}{P}\bigr)^\dagger\, 
{C}{P}\, Q_i^{}\, 
\bigl({C}{P}\bigr)^\dagger 
{C}{P} \Bigl| \bar{B}^0 \Bigr\rangle    
\nonumber \\  
&=&
\eta_B^* \eta_Q^{}\, (-1)^{l+1}\,
\Bigl\langle \bigl(\pi^+\pi^-\bigr)_l\, \pi_l^0 \Bigr| 
Q_i^\dagger\, \Bigl| B^0 \Bigr\rangle   \,\,,
\end{eqnarray}     
where    
$\,\bm{p}_-^{}=\bm{p},\,$  $\,\bm{p}_+^{}=-\bm{p},\,$  and  
$\,\bm{p}_0^{}=\bm{p}'.\,$     
The determined CP properties are independent of frame, so that 
we conclude, more generally, 
that\footnote{We can check the CP assignments by 
determining the CP of associated, two-body modes. Noting 
$\,CP|\rho^0\rangle = +|\rho^0\rangle\,$  and 
$\,CP|\rho^+\rangle = \eta_\rho|\rho^-\rangle,\,$ we find 
$\,\langle \pi^0 (p_0) \rho^0 (p_\rho)  | Q_i | \bar B^0\rangle 
 = \eta_B^\ast \eta_Q 
\langle \pi^0 (p_0) \rho^0 (p_\rho) | Q_i^\dagger | B^0\rangle,\, $ as well as
$\,\langle \pi^+ (p_+) \pi^- (p_-) | {\cal H}_{\rm str} 
| \rho^0 (p_\rho) \rangle = \langle \pi^+ (p_+) \pi^- (p_-) | 
{\cal H}_{\rm str}^\dagger | \rho^0  (p_\rho) \rangle,\,$  
where  $\,p_\rho = p_+ + p_-$ and $\bm{p}_\rho=\bm{0}.\,$ 
Moreover, 
$\,\langle \pi^+ (p_+) \rho^-(p_\rho) | Q_i | \bar B^0\rangle 
 = - \eta_B^\ast \eta_Q \eta_\rho^* \eta_\pi
\langle \pi^-(p_+) \rho^+ (p_\rho) | Q_i^\dagger | B^0\rangle\,$
and
$\,\langle \pi^- (p_-) \pi^0 (p_0) | {\cal H}_{\rm str} 
| \rho^- (p_\rho) \rangle = \eta_\rho \eta_\pi^*
\langle \pi^+ (p_-) \pi^0 (p_0) | 
{\cal H}_{\rm str}^\dagger | \rho^+  (p_\rho) \rangle,\,$  
where  $\,p_\rho = p_- + p_0\,$ and  $\,\bm{p}_\rho=\bm{0}.\,$     
Of course  $\,{\cal H}_{\rm str}={\cal H}_{\rm str}^\dagger.\,$     
Removing the matrix elements associated with $\rho\to \pi\pi$, 
we see that  $B\to \rho\pi$ transformation properties 
extracted from Eq.~(\ref{M_+-^l'}) are in accord with our
direct calculation. Choosing, in addition, $\,\eta_B^\ast \eta_Q =+1\,$ 
and  $\,\eta_\rho^*\eta_\pi=-1\,$  yields 
Eq.~(\ref{convdef}).}
\begin{eqnarray}   \label{M_+-^l'} 
\begin{array}{c}   \displaystyle   
\bar{\cal M}_{-+}^{l}\bigl(p_+^{},p_-^{},p_0^{}\bigr)  \,=\,  
-\eta_B^* \eta_Q^{} \sum_i^{} \lambda_i^{}\, \Bigl\langle 
\bigl[ \pi^+\bigl(p_-^{}\bigr)\, \pi^0\bigl(p_0^{}\bigr) \bigr]_l\, 
\pi^-\bigl(p_+^{}\bigr)_l \Bigr| 
Q_i^\dagger\, \Bigl| B^0 \Bigr\rangle   \,\,,  
\vspace{2ex} \\   \displaystyle   
\bar{\cal M}_{+-}^{l}\bigl(p_+^{},p_-^{},p_0^{}\bigr)   \,=\,  
-\eta_B^* \eta_Q^{} \sum_i^{} \lambda_i^{}\, \Bigl\langle 
\bigl[ \pi^-\bigl(p_+^{}\bigr)\, \pi^0\bigl(p_0^{}\bigr) \bigr]_l\, 
\pi^+\bigl(p_-^{}\bigr)_l \Bigr| 
Q_i^\dagger\, \Bigl| B^0 \Bigr\rangle   \,\,,  
\vspace{2ex} \\   \displaystyle   
\bar{\cal M}_{00}^{l}\bigl(p_+^{},p_-^{},p_0^{}\bigr)  \,=\,  
\eta_B^* \eta_Q^{}\, (-1)^{l+1}\, \sum_i^{} \lambda_i^{}\, 
\Bigl\langle \bigl(\pi^+(p_+)\pi^-(p_-)\bigr)_l\, \pi^0(p_0)_l \Bigr| 
Q_i^\dagger\, \Bigl| B^0 \Bigr\rangle   \,\,,
\end{array}     
\end{eqnarray}     
or that
\begin{eqnarray}   \label{Mcptrans}
\begin{array}{c}   \displaystyle   
\bar{\cal M}_{-+}^{l}\bigl(p_+^{},p_-^{},p_0^{}\bigr)  
\,\stackrel{\,CP}{\longleftrightarrow}\,
- \eta_B^* \eta_Q 
{\cal M}_{+-}^{l}\bigl(p_-^{},p_+^{},p_0^{}\bigr)   \,\,,  
\vspace{2ex} \\   \displaystyle   
\bar{\cal M}_{+-}^{l}\bigl(p_+^{},p_-^{},p_0^{}\bigr)   
\,\stackrel{\,CP}{\longleftrightarrow}\,
- \eta_B^* \eta_Q 
 {\cal M}_{-+}^{l}\bigl(p_-^{},p_+^{},p_0^{}\bigr)   \,\,,  
\vspace{2ex} \\   \displaystyle   
\bar{\cal M}_{00}^{l}\bigl(p_+^{},p_-^{},p_0^{}\bigr)  
\,\stackrel{\,CP}{\longleftrightarrow}\,
- \eta_B^* \eta_Q 
 {\cal M}_{00}^{l}\bigl(p_-^{},p_+^{},p_0^{}\bigr)   \,\,, 
\end{array}     
\end{eqnarray}     
where the $(-1)^l$ in the last line of Eq.~(\ref{M_+-^l'}) has been absorbed 
through the exchange  $\,p_+^{}\leftrightarrow p_-^{}.\,$  Thus we observe
that the CP transformation is tied to the mirror transformation
of the Dalitz plot,  $\,p_+^{}\leftrightarrow p_-^{}.\,$\footnote{We can use 
the amplitudes of Sec.~\ref{dcpva} to check Eq.~(\ref{Mcptrans}). 
Choosing  $\,\eta_B^*\eta_Q=+1\,$  and noting 
$\,f_\pm^{} \,\stackrel{\,p_+ \leftrightarrow p_-}{\longleftrightarrow}\,  
- f_\mp^{}\,$ 
and  
$\,f_0^{} \,\stackrel{\,p_+\leftrightarrow p_-}{\longleftrightarrow}\, - f_0^{},\,$    
so that    
$\,{\cal M}_{3\pi}^{}\bigl(p_+^{},p_-^{},p_0^{}\bigr) = 
a_{+-}^{}f_+^{} + a_{-+}^{}f_-^{} + a_{00}^{}f_0^{}\,$   
does yield 
$\,\bar{\cal M}_{3\pi}^{}\bigl(p_+^{},p_-^{},p_0^{}\bigr) = 
\bar a_{+-}^{}f_+^{} + \bar a_{-+}^{}f_-^{} + \bar a_{00}^{}f_0^{}\,$  
upon CP transformation, as per Eq.~(\ref{M_3pi}). 
}
Generalizing the definitions of Eq.~(\ref{aba}), we introduce 
\begin{eqnarray}   \label{M_gu} 
{\cal M}_g^{l}\bigl(p_+^{},p_-^{},p_0^{}\bigr) \! &=& \!
\mbox{$\frac{1}{2}$} \left[ \vphantom{|_|^|}
{\cal M}_{+-}^{l}\bigl(p_+^{},p_-^{},p_0^{}\bigr)  
+ {\cal M}_{-+}^{l}\bigl(p_+^{},p_-^{},p_0^{}\bigr) \right. 
\nonumber \\ && \left. \hspace{1em} \vphantom{|_|^|}
+\, (-1)^l {\cal M}_{+-}^{l}\bigl(p_-^{},p_+^{},p_0^{}\bigr)  
+ (-1)^l {\cal M}_{-+}^{l}\bigl(p_-^{},p_+^{},p_0^{}\bigr) 
\right] \,\,, 
\nonumber \\ 
{\cal M}_u^{l}\bigl(p_+^{},p_-^{},p_0^{}\bigr) \! &=& \!
\mbox{$\frac{1}{2}$} \left[ \vphantom{|_|^|} 
{\cal M}_{+-}^{l}\bigl(p_+^{},p_-^{},p_0^{}\bigr)  
+ {\cal M}_{-+}^{l}\bigl(p_+^{},p_-^{},p_0^{}\bigr)  
\right. \nonumber \\ && \left. \hspace{1em} \vphantom{|_|^|}
- (-1)^l {\cal M}_{+-}^{l}\bigl(p_-^{},p_+^{},p_0^{}\bigr)  
- (-1)^l
 {\cal M}_{-+}^{l}\bigl(p_-^{},p_+^{},p_0^{}\bigr)\right] \,\,,  \nonumber \\
{\cal M}_n^{l}\bigl(p_+^{},p_-^{},p_0^{}\bigr)  \! &=& \!
{\cal M}_{00}^{l}\bigl(p_+^{},p_-^{},p_0^{}\bigr)   \,\,,   
\end{eqnarray}
and
\begin{eqnarray}   \label{M_gubar} 
\bar{\cal M}_g^{l}\bigl(p_+^{},p_-^{},p_0^{}\bigr) \! &=& \!
\mbox{$\frac{1}{2}$} \left[ \vphantom{|_|^|} 
\bar{\cal M}_{+-}^{l}\bigl(p_+^{},p_-^{},p_0^{}\bigr)  
+ \bar{\cal M}_{-+}^{l}\bigl(p_+^{},p_-^{},p_0^{}\bigr)  
\right. \nonumber \\ && \left. \hspace{1em} \vphantom{|_|^|}
+ (-1)^l \bar{\cal M}_{+-}^{l}\bigl(p_-^{},p_+^{},p_0^{}\bigr)  
+ (-1)^l \bar{\cal M}_{-+}^{l}\bigl(p_-^{},p_+^{},p_0^{}\bigr)\right] 
\,\,,  \nonumber \\
\bar{\cal M}_u^{l}\bigl(p_+^{},p_-^{},p_0^{}\bigr) \! &=& \!
\mbox{$\frac{1}{2}$} \left[ \vphantom{|_|^|} 
\bar{\cal M}_{+-}^{l}\bigl(p_+^{},p_-^{},p_0^{}\bigr)  
+ \bar{\cal M}_{-+}^{l}\bigl(p_+^{},p_-^{},p_0^{}\bigr)  
\right. \nonumber \\ && \left. \hspace{1em} \vphantom{|_|^|}
- (-1)^l \bar{\cal M}_{+-}^{l}\bigl(p_-^{},p_+^{},p_0^{}\bigr)  
- (-1)^l \bar{\cal M}_{-+}^{l}\bigl(p_-^{},p_+^{},p_0^{}\bigr)\right] 
\,\,,  \nonumber \\
\bar{\cal M}_n^{l}\bigl(p_+^{},p_-^{},p_0^{}\bigr) \! &=& \!
\bar{\cal M}_{00}^{l}\bigl(p_+^{},p_-^{},p_0^{}\bigr)   \,\,,
\end{eqnarray}       
where we note\footnote{We can use the amplitudes
of Sec.~\ref{dcpva} to check Eq.~(\ref{CP[a_cdn]gen}) as well. 
Choosing  $\,\eta_B^*\eta_Q=+1,\,$  we see
$\,f_g^{} \,\stackrel{\,p_+\leftrightarrow p_-}{\longleftrightarrow}\, -f_g^{}\,$ 
and   
$\,f_u^{} \,\stackrel{\,p_+\leftrightarrow p_-}{\longleftrightarrow}\, +f_u^{},\,$ 
so that 
$\,{\cal M}_{3\pi}^{}\bigl(p_+^{},p_-^{},p_0^{}\bigr) = 
a_{g}^{}f_g^{} + a_{u}^{}f_u^{} + a_{n}^{}f_n^{}\,$  
does yield 
$\,\bar{\cal M}_{3\pi}^{}\bigl(p_+^{},p_-^{},p_0^{}\bigr) = 
\bar a_{g}^{}f_g^{} + \bar a_{u}^{}f_u^{} + \bar a_{n}^{}f_n^{}\,$   
under CP transformation, as per Eq.~(\ref{M_3pi}). 
}
\begin{eqnarray}   \label{CP[a_cdn]gen}
\begin{array}{c}   \displaystyle   
\bar {\cal M}_g^{l}\bigl(p_+^{},p_-^{},p_0^{}\bigr) 
\,\stackrel{\,CP}{\longleftrightarrow}\,
\,- \eta_B^* \eta_Q {\cal M}_g^{l}\bigl(p_-^{},p_+^{},p_0^{}\bigr) \,\,,  
\hspace{2em} 
\bar {\cal M}_u^{l}\bigl(p_+^{},p_-^{},p_0^{}\bigr) 
\,\stackrel{\,CP}{\longleftrightarrow}\,
- \eta_B^* \eta_Q {\cal M}_u^{l}\bigl(p_-^{},p_+^{},p_0^{}\bigr) \,\,,  
\vspace{2ex} \\   \displaystyle   
\bar {\cal M}_n^{l}\bigl(p_+^{},p_-^{},p_0^{}\bigr) 
\,\stackrel{\,CP}{\longleftrightarrow}
\, - \eta_B^* \eta_Q  {\cal M}_n^{l}\bigl(p_-^{},p_+^{},p_0^{}\bigr)\,\,.   
\end{array}
\end{eqnarray}     
To exploit these transformation properties, we parametrize
the $B\to \pi^+ \pi^-\pi^0$ amplitude as 
\begin{eqnarray}   \label{param}
{\cal M}_{3\pi}\bigl(p_+^{},p_-^{},p_0^{}\bigr)\,=\,\sum_{l=0}^\infty \left( 
{\cal M}_{+-}^{l} + {\cal M}_{-+}^{l} + {\cal M}_{00}^{l} \right)   
\,=\,   
\sum_{l=0}^\infty \left( 
{\cal M}_g^{l} + {\cal M}_u^{l} + {\cal M}_n^{l} \right)   \,\,,  
\end{eqnarray}     
where we have suppressed the common arguments throughout, so that
${\cal M}_{\pm\mp}^{l}$  is, indeed,   
${\cal M}_{\pm\mp}^{l}\bigl(p_+^{},p_-^{},p_0^{}\bigr)$.
Analogously, for the  $\,\bar B\to \pi^+ \pi^-\pi^0\,$  amplitude we have
\begin{eqnarray}  
\begin{array}{c}   \displaystyle   
\bar{\cal M}_{3\pi}\bigl(p_+^{},p_-^{},p_0^{}\bigr)\,=\,\sum_{l=0}^\infty \left( 
\bar{\cal M}_{+-}^{l} + \bar{\cal M}_{-+}^{l} + 
\bar{\cal M}_{00}^{l} \right)   
\,=\, \sum_{l=0}^\infty \left( 
\bar{\cal M}_g^{l} + \bar{\cal M}_u^{l} + \bar{\cal M}_n^{l} \right)   \,\,.  
\end{array}     
\end{eqnarray}     
A two-body decay mechanism distinguishes the terms of fixed
$l$; however, our parametrization is not limited by this 
assumption. Note that ${\cal M}_{+-}^l\bigl(p_+^{},p_-^{},p_0^{}\bigr)$, 
e.g., includes the product  $a_{+-}^{}f_+^{}$  of Sec.~\ref{dcpva}.
With these definitions, the untagged decay rate can be written as  
\begin{equation}
\label{m3pigen}
\bigl|{\cal M}_{3\pi}^{}\bigr|^2  + \bigl|\bar{\cal M}_{3\pi}^{}\bigr|^2 
\,=\, 
\sum_{\kappa \in g,u,n} \sum_{j,l=0}^{\infty} \left[ \left( 
{\cal M}_\kappa^{j\,*} {\cal M}_\kappa^{l} 
+ \bar {\cal M}_\kappa^{j\,*} \bar {\cal M}_\kappa^{l} \right) 
+ 2 \sum_{\lambda > \kappa }
\mbox{Re}({\cal M}_\kappa^{j\,*} {\cal M}_\lambda^{l} + 
\bar {\cal M}_\kappa^{j\,*} \bar {\cal M}_\lambda^{l})
\right] \,\,,
\end{equation}
where $\lambda \in g,u,n$ and $\lambda > \kappa$ 
means that the $g,u,n$ subscripts are not repeated. 
Applying Eq.~(\ref{CP[a_cdn]gen}), we observe that 
\begin{equation}\label{realdeal}
\bigl|{\cal M}_{3\pi}^{}\bigl(p_+^{},p_-^{},p_0^{}\bigr)\bigr|^2  + 
\bigl|\bar {\cal M}_{3\pi}^{}\bigl(p_+^{},p_-^{},p_0^{}\bigr)\bigr|^2 
 \,\stackrel{\,CP}{\longleftrightarrow}\,  
\bigl|{\cal M}_{3\pi}^{}\bigl(p_-^{},p_+^{},p_0^{}\bigr)\bigr|^2  + 
\bigl|\bar {\cal M}_{3\pi}^{}\bigl(p_-^{},p_+^{},p_0^{}\bigr)\bigr|^2  \,\,,
\end{equation}
so that the untagged decay width, $\Gamma_{3\pi}$, realized from 
integrating Eq.~(\ref{m3pigen}) over the invariant phase
space, as in Eq.~(\ref{Gamma_3pi}), is a manifestly CP-even quantity.
This follows because the integration region is 
itself $p_+\leftrightarrow p_-$ symmetric. 
The population asymmetry, however, allows us to observe direct CP violation. 
Forming 
\begin{eqnarray}  
{\cal A}_{3\pi}^{}  \,\equiv\,  
{ \Gamma_{3\pi}^{} \bigl[s_{+0}^{}\mbox{$>$}s_{-0}^{}\bigr] 
 - \Gamma_{3\pi}^{} \bigl[s_{+0}^{}\mbox{$<$}s_{-0}^{}\bigr]  
 \over 
 \Gamma_{3\pi}^{} \bigl[s_{+0}^{}\mbox{$>$}s_{-0}^{}\bigr] 
 + \Gamma_{3\pi}^{} \bigl[s_{+0}^{}\mbox{$<$}s_{-0}^{}\bigr] }  \,\,,
\end{eqnarray}     
where 
\begin{eqnarray}    
\Gamma_{3\pi}^{} \bigl[s_{+0}^{}\mbox{$>$}s_{-0}^{}\bigr] 
 - \Gamma_{3\pi}^{} \bigl[s_{+0}^{}\mbox{$<$}s_{-0}^{}\bigr]  
\,=\,  
\int_{s_{+0}^{}>s_{-0}^{}}\!\!\!\!\!\! &&{\rm d}\Phi\; 
\left\{
\left[ \vphantom{|_|^|}
\bigl|{\cal M}_{3\pi}^{}\bigl(p_+^{},p_-^{},p_0^{}\bigr)\bigr|^2  + 
\bigl|\bar{\cal M}_{3\pi}^{}\bigl(p_+^{},p_-^{},p_0^{}\bigr)\bigr|^2 
\right]  \right.
\nonumber \\
&& \hspace*{2em}  
- \left. \left[ \vphantom{|_|^|}
\bigl|{\cal M}_{3\pi}^{}\bigl(p_-^{},p_+^{},p_0^{}\bigr) \bigr|^2 + 
\bigl|\bar{\cal M}_{3\pi}^{}\bigl(p_-^{},p_+^{},p_0^{}\bigr)\bigr|^2
\right] \right\}   \,\,,  \hspace*{2em}
\end{eqnarray}     
and noting Eq.~(\ref{realdeal}), we observe that ${\cal A}_{3\pi}^{}$
is odd under  $\,p_+^{}\leftrightarrow p_-^{}\,$  exchange and that it is also 
manifestly CP odd. Consequently, the observation of a population
asymmetry, that is, 
the failure of mirror symmetry across the Dalitz plot of the
untagged decay rate,  
signals the presence of direct CP violation. 
To determine the amplitude combinations which give rise to a
population asymmetry, 
let us consider the 
transformation properties of the amplitudes of Eq.~(\ref{m3pigen})
under $p_+ \leftrightarrow p_-$, as this is the mirror transformation
of the Dalitz plot. We have 
\begin{eqnarray}   \label{mirror}
{\cal M}_{g}^{l} \,
\stackrel{\,p_+ \leftrightarrow p_-}{\longleftrightarrow}\,
(-1)^{l} {\cal M}_{g}^{l} \,\,,  
\hspace{2em}  
{\cal M}_u^{l} \,\stackrel{\,\,p_+ \leftrightarrow p_-}{\longleftrightarrow}\,
-(-1)^{l}\bar {\cal M}_u^{l} \,\,,  
\hspace{2em}  
{\cal M}_{n}^{l} \,
\stackrel{\,p_+ \leftrightarrow p_-}{\longleftrightarrow}\,
(-1)^{l} {\cal M}_{n}^{l} \,\,,  
\end{eqnarray}   
where analogous relationships hold for the barred amplitudes as well. 
We thus determine that if $j+l$ is even only the amplitude combinations
\begin{eqnarray}  \label{gen:evenl}
{\cal M}_g^{j\,*} {\cal M}_u^{l}+\bar {\cal M}_g^{j\,*} \bar {\cal M}_u^{l}  
\hspace{2em} {\rm and} \hspace{2em}   
{\cal M}_u^{j\,*} {\cal M}_n^{l}+\bar {\cal M}_u^{j\,*} 
\bar {\cal M}_n^{l}   \,\,,
\end{eqnarray}     
are odd under the mirror symmetry of the Dalitz
plot, whereas if $j+l$ is odd only 
\begin{eqnarray}  \label{gen:oddl}
&&{\cal M}_g^{j\,*} {\cal M}_g^{l}+\bar {\cal M}_g^{j\,*} 
\bar {\cal M}_g^{l}\,\,,  \nonumber\\
&&{\cal M}_u^{j\,*} {\cal M}_u^{l}+\bar {\cal M}_u^{j\,*} 
\bar {\cal M}_u^{l}\,\,,  \nonumber\\
&&{\cal M}_n^{j\,*} {\cal M}_n^{l}+\bar {\cal M}_n^{j\,*} 
\bar {\cal M}_n^{l}\,\,,  
\hspace{2em} {\rm and} \hspace{2em}   
{\cal M}_g^{j\,*} {\cal M}_n^{l}+\bar {\cal M}_g^{j\,*} 
\bar {\cal M}_n^{l}\,\,,
\end{eqnarray}     
are odd under the mirror symmetry of the Dalitz plot. These
amplitude combinations generate a population asymmetry in the untagged
decay rate, which we have established as a signal of direct
CP violation. 
As a plurality of amplitude combinations exist, they 
can act in concert to dilute the population asymmetry we have discussed. 
Nevertheless,
the observation of a population asymmetry in the untagged decay rate
to three pseudoscalar mesons is an unambiguous signal
of direct CP violation. 

Our discussion generalizes
the CP-odd amplitudes found under an assumption of $\rho$ 
dominance, Eq.~(\ref{ac*ad}), in two ways. 
Firstly, Eq.~(\ref{gen:evenl}) shows that combinations 
analogous in structure to Eq.~(\ref{ac*ad}) exist
for all even $j+l$. Equation (\ref{gen:oddl}) shows
that 
direct CP violation can also be generated for odd $j+l$, 
such as through $s$-wave and $p$-wave interference. 

\section{Comparison to the partial-rate asymmetry}
\label{cf}

We wish to compare the size of the population asymmetry 
in the untagged decay rate with that of the partial rate
asymmetry in the same decay channel. 
We will show that 
the two asymmetries can be comparable in size, favoring 
the study of direct CP-violation in the
untagged decay rate on statistical grounds. 
We begin by considering the amplitude combination
$\,a_g^* a_u^{}+\bar a_g^* \bar a_u^{}\,$.  
We can write $a_{kl}$ of Eq.~(\ref{M_pirho})
in terms of its tree and penguin contributions, as per
$\exp({{\rm i}\beta}) \,a_{kl}^{}\equiv T_{kl}^{}\, 
\exp({-{\rm i}\alpha}) + P_{kl}^{}\,$, 
where both $T_{kl}^{}$  and  $P_{kl}^{}$ are complex, 
and  the weak phases $\beta$  and  $\gamma$ are given by 
$\,\exp(-{\rm i}\beta)\equiv V_{tb}^{}V_{td}^*/
|V_{tb}^{}V_{td}^*|\,$  
and
$\,\exp({\rm i}\gamma)= V_{ub}^* V_{ud}^{}/
|V_{ub}^*V_{ud}^{}|\,$. 
We take   $V_{kl}^{}$ to be an element of the CKM matrix, and 
we employ CKM unitarity 
to write $\alpha + \beta+\gamma = \pi\,$. 
Noting $\exp(-{\rm i}\beta) \,\bar a_{kl}^{}
=T_{kl}^{}\, \exp({\rm i}\alpha) + P_{kl}^{}\,$, 
we have      
\begin{eqnarray}   \label{a_cd}
\begin{array}{c}   \displaystyle      
a_g^{}  \,=\,  
T_g^{}\, {\rm e}^{-{\rm i}\alpha}+P_g^{}   \,\,,  
\hspace{2em}  
a_u^{}  \,=\,  
T_u^{}\, {\rm e}^{-{\rm i}\alpha}+P_u^{}   \,\,,  
\vspace{1ex} \\   \displaystyle   
\bar a_g^{}  \,=\,  
T_g^{}\, {\rm e}^{{\rm i}\alpha}+P_g^{}   \,\,,  
\hspace{2em}  
\bar a_u^{}  \,=\,  
-T_u^{}\, {\rm e}^{{\rm i}\alpha}-P_u^{}   \,\,,    
\end{array}   
\end{eqnarray}   
where  $\,T_g^{}=T_{+-}^{}+T_{-+}^{}\,$  and  
$\,T_u^{}=T_{+-}^{}-T_{-+}^{},\,$  with similar relations for $P_{g,u}^{}$.  
We omit the overall factors of $\exp(\pm {\rm i} \beta)$ which ought to 
appear in Eq.~(\ref{a_cd}) as they contribute to neither 
$|{\cal M}_f|^2$, $|\bar {\cal M}_f|^2$, 
nor  $q {\cal M}_f^* \bar {\cal M}_f^{} /p$ --- we note that 
$q/p = \exp(-2{\rm i} \beta)$ in the SM 
if $|q/p|=1$. With 
$\, r_\kappa^{} \, \exp({\rm i}\delta_\kappa^{})
\equiv P_\kappa^{}/T_\kappa^{}\,$   
for  $\,\kappa=g,u\,$, so that  
$\,r_\kappa^{}= \bigl|P_\kappa^{}/T_\kappa^{}\bigr|,\,$  
we find~\cite{gardner} 
\begin{eqnarray}   \label{aa+baba}
a_g^{}\, a_u^{*} + \bar a_g^{}\, \bar a_u^{*}  \,=\,  
-2\, T_g^{}\, T_u^{*}\, \sin\alpha   
\Bigl[ r_g^{}  \, \sin\delta_g^{} 
      + r_u^{} \, \sin\delta_u^{} 
      - {\rm i}\, 
\bigl( r_g^{}  \, \cos\delta_g^{} 
      -  r_u^{} \, \cos\delta_u^{} \bigr) \Bigr] \, 
\;. 
\end{eqnarray}   
The decay width associated with this amplitude combination, 
$\Gamma_{3\pi}^{(2)}$ in Eq.~(\ref{Gamma23}), contains
\begin{eqnarray}     
{\rm Re} \left[ 
\left( a_g^{} a_u^{*}+\bar a_g^{} \bar a_u^{*} \right) f_g^{} f_u^{*} 
\right] \,\,,
\end{eqnarray}     
so that 
it is sensitive to both its 
real and imaginary parts, because the function $f_g^{} f_u^{*}$ 
is also complex 
--- its imaginary part is generated by the nonzero $\rho$-resonance width. 
If we assume $T_g^{}\,T_u^*$ to be real, for definiteness, and consider
the imaginary part of Eq.~(\ref{aa+baba}), 
which accompanies ${\rm Im}(f_g^{} f_u^*)\,$, we see
that $\Gamma_{3\pi}^{(2)}$ can be nonzero, so that 
direct CP violation can exist, even if the strong phases of $a_i$ were
to vanish. Merely $r_g$ or $r_u$ must be nonzero, which is easily
satisfied. Consequently the observables of Eq.~(\ref{asymobs}) can be
nonzero irrespective of the strong phases of $a_i$ if the 
width of the resonance in the CP-enantiomer
is nonzero. Note that were 
$T_g T_u^*$ complex, 
the $\rho$ width would not be needed to 
generate DCPV. 
Similar considerations exist in $B\to V_1 V_2$ decays. 
There, too, direct CP violation can occur in the untagged decay rate;
moreover, its presence does not demand 
the strong phase of the weak transition amplitude to be 
nonzero~\cite{valencia}.

In contrast, the partial-rate asymmetry 
associated with $B^0 \to \rho^+ \pi^-$ decay is  
\begin{eqnarray}   \label{A_rate} 
\Gamma_{B^0\to\rho^+\pi^-}^{} - \Gamma_{\bar B^0\to\rho^-\pi^+}^{} 
\,\propto\,   
\bigl| a_{+-}^{} \bigr| ^2 - \bigl| \bar a_{-+}^{} \bigr| ^2 
\,=\,  
-4 \left| T_{+-}^{} 
\right|^2 r_{+-}^{}\, \sin\delta_{+-}^{}\, \sin\alpha  \,\,,  
\end{eqnarray}     
so that both $\alpha$ and $\delta_{+-}^{}$ must be nonzero to yield
a nonzero partial rate asymmetry. Generally one expects
$|{\cal M}_{\rho^+\pi^-}| > 
|{\cal M}_{\rho^-
\pi^+}|$~\cite{AliKL,Cheng:1999xj,qcdf,BenNeu}; 
this is borne out by experiment~\cite{babar628}. In the limit where 
${\cal M}_{\rho^-\pi^+}$ 
becomes negligibly small, we see that
Eq.~(\ref{aa+baba}) becomes
\begin{eqnarray}   
a_g^{}\, a_u^{*} + \bar a_g^{}\, \bar a_u^{*}  \,\to\,  
-4 \left| T_{+-}^{} 
\right|^2 r_{+-}^{}\, \sin\delta_{+-}^{}\, \sin\alpha  \,\,,  
\end{eqnarray}   
so that the population asymmetry of Eq.~(\ref{asymobs}) is identical
to the partial rate asymmetry in this limit.

In the following section, we illustrate these ideas numerically 
and discuss their experimental realization in greater detail. 
After our analysis of $B\to \rho\pi \to \pi^+ \pi^- \pi^0$ decay, we
turn to other modes, studying
$\,B\to D^{*\pm}D^\mp\to D^+D^-\pi^0,\,$  
$\,B_s^{}\to K^{*\pm}K^\mp\to K^+K^-\pi^0,\,$  and  
$\,B_s^{}\to D_s^{*\pm}D_s^\mp\to D_s^+D_s^-\pi^0\,$  decays.
We discuss the utility of these modes in realizing tests of the SM.

\section{Numerical Estimates\label{results}}    
     
In this section we estimate the population asymmetries associated
with the untagged decay rate in a variety of decays to 
self-conjugate
final states. We do this for definiteness, though our essential 
conclusions do not rely on these estimates: the population asymmetry
in the untagged decay rate can be of the same numerical size 
as the partial  rate
asymmetry in the comparable, tagged decay, making the search
for DCPV in the untagged process advantageous. 

We begin our numerical analysis by 
considering the effective, 
weak Hamiltonian for the decay  $\,b\to q q'\bar q'\,$   
and its CP-conjugate,  where  $\,q=d,s\,$  and  
$\,q'=u,d,s,c.\,$    
In the SM, we have~\cite{BucBL}      
\begin{eqnarray}   \label{H_eff}    
{\cal H}_{\rm eff}^{}  \,=\,  
{G_{\rm F}^{}\over\sqrt 2} \left[ 
\lambda_u^{(q)} \left( C_1^{} Q_1^{u} + C_2^{} Q_2^{u} \right)   
+ \lambda_c^{(q)} \left( C_1^{} Q_1^{c} + C_2^{} Q_2^{c} \right)   
- \lambda_t^{(q)} \left( 
\sum_{i=3}^{10} C_i^{} Q_i^{} 
+ C_{8G}^{} Q_{8G}^{} \right)   
\right]   
+ {\rm H.c.}  \,\,,  \hspace*{1em}
\end{eqnarray}     
where $G_{\rm F}^{}$  is the Fermi coupling constant and  the factor 
$\,\lambda_{q'}^{(q)}\equiv V_{q'b}^{} V_{q'q}^{*}\,$ contains a 
CKM matrix element $V_{ij}$. 
The Wilson coefficients $C_i^{}$  are 
evaluated at the renormalization scale  $\mu$, and  
$Q_{1,\cdots,10}^{}$ are four-quark operators, whereas  
$Q_{8G}^{}$  is the chromomagnetic penguin operator.  
The $C_i^{}$ and $Q_i^{}$ are detailed 
in Ref.~\cite{BucBL}, though we interchange  
$\,C_1^{} Q_1^q\leftrightarrow C_2^{} Q_2^q\,$, so that  $\,C_1^{}\sim 1\,$  
and  $\,C_1^{}>\bigl|C_2^{}\bigr|\gg C_{3,\dots,10,8G}\,$.
For the CKM matrix elements, we adopt the Wolfenstein 
parametrization~\cite{wolfenstein}, 
\begin{eqnarray}    
\begin{array}{c}   \displaystyle   
V_{ud}^{}  \,=\,  1-\lambda^2/2   \,\,,   \hspace{2em}  
V_{us}^{}  \,=\,  \lambda   \,\,,   \hspace{2em}  
V_{ub}^{}  \,=\,  A \lambda^3\, (\rho-{\rm i}\eta)   \,\,,  
\vspace{2ex} \\   \displaystyle   
V_{cd}^{}  \,=\,  -\lambda   \,\,,   \hspace{2em}  
V_{cs}^{}  \,=\,  1-\lambda^2/2   \,\,,   \hspace{2em}  
V_{cb}^{}  \,=\,  A \lambda^2   \,\,,  
\vspace{2ex} \\   \displaystyle   
V_{td}^{}  \,=\,  A \lambda^3\, (1-\rho-{\rm i}\eta)   \,\,,   \hspace{2em}  
V_{ts}^{}  \,=\,  -A\lambda^2   \,\,,   \hspace{2em}  
V_{tb}^{}  \,=\,  1  \,\,,   
\end{array}      
\end{eqnarray}      
neglecting  ${\cal O}\bigl(\lambda^4\bigr)$,  
where the parameters are given 
by~\cite{AliKL}
\begin{eqnarray}   \label{ckmfit}
\lambda  \,=\,  0.2205   \,\,,  \hspace{2em}  
A  \,=\,  0.81   \,\,,  \hspace{2em}  
\rho  \,=\,  0.12   \,\,,  \hspace{2em}  
\eta  \,=\,  0.34   \,\,.  
\end{eqnarray}     

The three-body decays we consider proceed from 
hadronic two-body weak transitions of the form  $\,B\to X Y,\,$    
for which the amplitude is given by  
\begin{eqnarray}   
{\cal M}_{B\to XY}^{}  \,=\, 
\langle X Y|{\cal H}_{\rm eff}^{}|B\rangle   \,\,.   
\end{eqnarray}     
For definiteness and simplicity, we employ the generalized 
factorization approximation~\cite{AliKL,gfa}, which makes use of 
next-to-leading-order perturbative-QCD to estimate 
the needed strong phases.  
In this approach, the matrix elements  
$\,C_i^{}(\mu)\, \langle X Y|Q_i^{}(\mu)|B\rangle\,$  
are approximated by  
$\,C_i^{\rm eff} \langle X Y|Q_i^{}|B\rangle_{\rm tree}^{},\,$   
where the $C_i^{\rm eff}$  are effective Wilson coefficients  
and  the $\langle X Y|Q_i^{}|B\rangle_{\rm tree}^{}$  
are calculated under the factorization assumption.     
The $C_i^{\rm eff}$  enter 
the decay amplitudes through the combinations  
\begin{eqnarray}   
a_{2i-1}^{}  \,=\,  
C_{2i-1}^{\rm eff} + {C_{2i}^{\rm eff}\over N_{\rm c}^{}}   \,\,, 
\hspace{2em}  
a_{2i}^{}  \,=\,   
C_{2i}^{\rm eff} + {C_{2i-1}^{\rm eff}\over N_{\rm c}^{}}   \,\,,  
\end{eqnarray}     
where  $\,i=1,\cdots,5\,$   and  
$\,N_{\rm c}^{}=3\,$. 
We adopt the  $a_i^{}$   
values at  $\,\mu=2.5\,\rm GeV\,$  calculated in Ref.~\cite{AliKL},  
collected in their Tables~VI and~VII.     
For reference, we report these numbers in Table~\ref{a_i} 
in the Appendix.  Our purpose is to gain an impression of the
relative size of the partial rate and 
population asymmetries in a variety of modes.

\subsection{\mbox{\boldmath$B\to\pi^+\pi^-\pi^0\,$}\label{3pi}}

We now evaluate the CP-odd observables in untagged  
$\,B^0,\bar B^0\to \pi^+\pi^-\pi^0\,$  decays. 
To begin, we assume, as in Sec.~\ref{dcpv}, that 
$\,B^0,\bar B^0\to \pi^+\pi^-\pi^0\,$  decays are dominated
by $\rho\pi$ intermediate states. The $\,B\to\rho\pi\,$  amplitudes
we employ, 
as well as our ancillary input parameters and form factors, are given
in the Appendix. 
The population asymmetries of Eq.~(\ref{asymobs}) are 
computed using Eqs.~(\ref{Gamma1},\ref{Gamma23}). We begin by
computing 
\begin{equation}
\Gamma_{ij}  \,\equiv\,   
\int {\rm d}\Phi\; 
f_i^* f_j^{}  \,\,,  
\end{equation}
to find 
\begin{eqnarray} \label{gamma_num}
\begin{array}{c}   \displaystyle   
\Gamma_{gu}^{} \bigl[s_{+0}^{}\mbox{$>$}s_{-0}^{}\bigr] = 
-\Gamma_{gu}^{} \bigl[s_{+0}^{}\mbox{$<$}s_{-0}^{}\bigr] = 
(-54.57 - 3.05\,{\rm i})\times 10^{10}\, \Gamma_{B^0}^{}   \,\,, 
\vspace{2ex} \\   \displaystyle   
\Gamma_{un}^{} \bigl[s_{+0}^{}\mbox{$>$}s_{-0}^{}\bigr] = 
-\Gamma_{un}^{} \bigl[s_{+0}^{}\mbox{$<$}s_{-0}^{}\bigr] = 
(2.01 - 0.08\,{\rm i})\times 10^{10}\, \Gamma_{B^0}^{}  \,\,,
\end{array}
\end{eqnarray}
where the arguments in brackets indicate the regions in phase
space over which the integral is calculated. The bracketed 
regions can also be selected by fixing the sign of $\cos \theta$, where
$\theta$ is the helicity angle. 
We define $\theta$ in 
$\,B(p_B) \to \pi^+ (p_+) \pi^- (p_-) \pi^0 (p_0)\,$ decay via
\begin{equation}
\cos \theta = 
\frac{\bm{p}_-^\prime \cdot \bm{p}_0^\prime}
{\bigl|\bm{p}_-^\prime\bigr| \, \bigl|\bm{p}_0^\prime\bigr|} \,\,, 
\end{equation}
where the primed variables refer to the momenta in the rest frame 
of the $\pi^+(p_+^\prime) \pi^- (p_-^\prime)$ pair, so that 
$\,\bm{p}_+^\prime + \bm{p}_-^\prime=0.\,$   
Note that  $\,s_{-0}^{}\stackrel{<}{\!{}_>}s_{+0}^{}\,$  
corresponds to  $\,\cos\theta\stackrel{<}{\!{}_>}0.\,$ 
The Im parts of $\Gamma_{gu}$ and $\Gamma_{un}$ 
are relatively small, so that short-distance strong phases
are necessary, from a practical viewpoint, to 
observe a population asymmetry. Employing the
model of Ref.~\cite{AliKL} and the parameters reported
in the Appendix, we find 
\begin{equation}
a_g^\ast  a_u^{} + \bar a_g^\ast \bar a_u^{}  \,=\,  
(0.616 - 1.683\,{\rm i})\times 10^{-18}   \,\,,   
\hspace{2em}  
a_u^\ast  a_n^{} + \bar a_u^\ast \bar a_n^{}  \,=\,   
(0.186 - 0.664\,{\rm i})\times 10^{-18}   \,\,.
\end{equation}
It is useful to report the separate 
$\Gamma^{(i)}$ contributions to the asymmetries, as per 
Eqs.~(\ref{Gamma1},\ref{Gamma23}), as well. To this end, we define
\begin{equation}
{\cal B}^{(i)} \equiv 
\frac{\Gamma^{(i)}_{B\to 3\pi} [ s_{+0} > s_{-0} ]}{\Gamma_{B^0}} 
\,\,, \hspace{3em} 
{\cal B}_{3\pi} \equiv 
\frac{\Gamma(B^0 , \bar B^0 \to \pi^+ \pi^- \pi^0)}{\Gamma_{B^0}}\,\,,
\end{equation}
so that Eq.~(\ref{asymobs}) can be rewritten 
\begin{equation}
{\cal A}_{3\pi}^{(i)} = 
\frac{2 {\cal B}^{(i)}}{{\cal B}_{3\pi}} \,\,.
\end{equation} 
We have 
\begin{eqnarray}   \label{cd}
\begin{array}{c}   \displaystyle   
{\cal B}_{3\pi}^{(2)} \bigl[s_{+0}^{}\mbox{$>$}s_{-0}^{}\bigr] 
\,=\,  -7.8\times 10^{-7}   \,\,,  
\hspace{2em}  
{\cal B}_{3\pi}^{(3)} \bigl[s_{+0}^{}\mbox{$>$}s_{-0}^{}\bigr] 
\,=\,  +6.4\times 10^{-9}   \,\,,  
\vspace{2ex} \\   \displaystyle   
{\cal B}_{3\pi}^{}  \,=\,  
2\, {\cal B}_{3\pi}^{(1)} \bigl[s_{+0}^{}\mbox{$>$}s_{-0}^{}\bigr] 
\,=\,  46\times 10^{-6}   \,\,, 
\end{array}     
\end{eqnarray}     
to yield 
\begin{equation} \label{A23pi}
{\cal A}_{3\pi}^{(2)} = -3.4\% \,\, , \hspace{3em} 
{\cal A}_{3\pi}^{(3)} = 0.03\% \,\,.
\end{equation}
The CP-violating asymmetry of  Eq.~(\ref{a_3pi})  is thus 
\begin{eqnarray}   \label{A_3pi}
{\cal A}_{3\pi}^{}  \,=\,  -3.3\%   \,\,, 
\end{eqnarray}     
which is clearly dominated by ${\cal A}_{3\pi}^{(2)}$, 
the  $\Gamma_{3\pi}^{(2)}$  contribution.  
This is expected because  $\Gamma_{3\pi}^{(3)}$  
contains, in comparison, the  
$B (\bar B) \to \rho^0\pi^0$  amplitude, which is 
``color-suppressed'' at tree-level; it is also 
suppressed by phase space, as per Eq.~(\ref{gamma_num}).

In constructing the population asymmetry, we need not integrate
over the entire, allowed region. We can also study the population
asymmetry over regions of the Dalitz plot; this is realized
in Table~\ref{G23}. Selecting the regions where the $\rho$ bands
overlap enhances the population asymmetry, but to the detriment
of its detectability, as the branching ratio into the selected
regions of the Dalitz plot, ${\cal B}_{3\pi}$, becomes considerably smaller. 
Indeed, for a direct CP search it would be more efficacious to use
the entire allowed region. 

\begin{table}[ht]   
\caption{\label{G23}%
The ratios 
$\,{\cal B}^{(i)}\equiv{\cal B}_{3\pi}^{(i)}
\bigl[s_{+0}^{}\mbox{$>$}s_{-0}^{}\bigr]=\Gamma_{3\pi}^{(i)}
\bigl[s_{+0}^{}\mbox{$>$}s_{-0}^{}\bigr]/\Gamma_{B^0}^{}\,$   
for  $\,i=2,3,\,$  total branching-ratio  
$\,{\cal B}_{3\pi}^{}=\Gamma_{3\pi}^{}/\Gamma_{B^0}^{},\,$  
and population asymmetry  ${\cal A}_{3\pi}^{}$,   
corresponding to various integration regions within the Dalitz plot, 
with  $\,m_{kl}^{}=\sqrt{s_{kl}^{}}.\,$    
The ${\cal B}_{3\pi}^{}$ entries includes contributions from both 
sides of the $\,s_{+0}^{}=s_{-0}^{}\,$ line.}  
\centering   \small 
\vskip 0.5\baselineskip
\begin{tabular}{c||c|c|c|c}%
\hline \hline  
Region within Dalitz plot  &  ${\cal B}^{(2)}$  &  
${\cal B}^{(3)}$    &  ${\cal B}_{3\pi}^{}$     &  
${\cal A}_{3\pi}^{}=\displaystyle   
\frac{ 2{\cal B}^{(2)}+2{\cal B}^{(3)}}{{\cal B}_{3\pi}^{}}\,\,\, 
(\%)  \vphantom{\Biggl]_|^|}$  
\\ \hline \hline && && \vspace{-2ex} 
\\   
entire allowed region$\vphantom{\big|}$  &  $-7.8\times 10^{-7}$  &  
$+6.4\times 10^{-9}$  &  $46\times 10^{-6}$  &  $-3.3$ 
\vspace{-2ex} \\ && && \\ \hline && && \vspace{-2ex} 
\\   
$m_\rho^{}-3\Gamma_\rho^{}<m_{+0}^{},m_{-0}^{}
<m_\rho^{}+3\Gamma_\rho^{}\; \vphantom{\big|}$  &  
$-6.6\times 10^{-8}$  &  $-4\times 10^{-12}$  &  
$22\times 10^{-7}$  &  $-6.1$   
\\    
$m_\rho^{}-2\Gamma_\rho^{}<m_{+0}^{},m_{-0}^{}
<m_\rho^{}+2\Gamma_\rho^{}\; \vphantom{\big|}$  &  
$-2.8\times 10^{-8}$  &  $-1\times 10^{-12}$  &  
$73\times 10^{-8}$  &  $-7.6$   
\\   
$m_\rho^{}-\Gamma_\rho^{}<m_{+0}^{},m_{-0}^{}
<m_\rho^{}+\Gamma_\rho^{}\; \vphantom{\big|}$  &  
$-1.4\times 10^{-9}$  &  $-4\times 10^{-14}$  &  
$31\times 10^{-9}$  &  $-9.3$  
\vspace{-2ex} \\ && && \\ \hline && && \vspace{-2ex} 
\\   
$m_\rho^{}-3\Gamma_\rho^{}<m_{+-}^{},m_{-0}^{}
<m_\rho^{}+3\Gamma_\rho^{}\; \vphantom{\big|}$  &  
$-3.8\times 10^{-8}$  &  $+0.4\times 10^{-8}$  &  
$27\times 10^{-7}$  &  $-2.6$  
\vspace{-2.3ex} \\ && && 
\\  
$m_\rho^{}-2\Gamma_\rho^{}<m_{+-}^{},m_{-0}^{}
<m_\rho^{}+2\Gamma_\rho^{}\; \vphantom{\big|}$  &  
$-1.5\times 10^{-8}$  &  $+0.2\times 10^{-8}$  &  
$10\times 10^{-7}$  &  $-2.3$  
\vspace{-2.3ex} \\ && && 
\\   
$m_\rho^{}-\Gamma_\rho^{}<m_{+-}^{},m_{-0}^{}
<m_\rho^{}+\Gamma_\rho^{}\; \vphantom{\big|}$  &  
$-1.0\times 10^{-9}$  &  $+0.3\times 10^{-9}$  &  
$76\times 10^{-9}$  &  $-1.9$  
\vspace{-2.3ex} \\ && && 
\\
\hline \hline  
\end{tabular}
\end{table}

More realistically, $B\to 3\pi$ decays will contain
contributions from other than $B\to \rho^i\pi^j\to 3\pi$. 
We can isolate the individual  $\Gamma_{3\pi}^{(2,3)}$  
contributions by restricting the integration regions to the  
$\rho$  bands,  which lie close to the edges of the Dalitz plot, 
as shown in Fig.~\ref{Bto3pi}.  
To realize this, we integrate over the region of $3\pi$ phase-space 
satisfying the requirement that the invariant mass of two of the three pions 
yield the $\rho$ mass within an interval of $M_\rho\pm \delta$.   
This amounts to calculating the effective width
\begin{eqnarray}   \label{Geff}
\begin{array}{c}   \displaystyle   
\Gamma_{B\to\rho(p_1+p_2)\pi(p_3)}^{\rm eff}    
\,=\,  
\Gamma\bigl(B\to \pi(p_1) \pi(p_2) \pi(p_3)\bigr) \Bigl|_{
(M_\rho^{}-\delta)^2\le s_{12}^{}\le (M_\rho^{}+\delta)^2}^{}   \,\,.  
\end{array}     
\end{eqnarray}     
In an experimental analysis the strength associated with the complete
$\rho$ line shape would be assessed; our effective width Ansatz
is a simple simulacrum of this more sophisticated analysis.
In order to avoid double counting in determining the population 
asymmetries, we further require that the 
$\rho^-\pi^+$  $\bigl(\rho^+\pi^-\bigr)$  contributions 
be restricted to the  $\,s_{+0}^{}>s_{-0}^{}\,$ 
$\bigl(s_{+0}^{}<s_{-0}^{}\bigr)$  region.   
The resulting effective branching ratios for  
$\,\delta=0.3\,\rm GeV\,$  are  
\begin{eqnarray}   \label{effB[rhopi]}
\begin{array}{c}   \displaystyle   
{\cal B}_{B^0\to\rho^-\pi^+}^{\rm eff} 
\bigl[s_{+0}^{}\mbox{$>$}s_{-0}^{}\bigr] 
\,=\,  4.9\times 10^{-6}   \,\,,  
\hspace{2em}   
{\cal B}_{B^0\to\rho^+\pi^-}^{\rm eff} 
\bigl[s_{+0}^{}\mbox{$<$}s_{-0}^{}\bigr]  
\,=\,  17.6\times 10^{-6}   \,\,,  
\vspace{1ex} \\   \displaystyle   
{\cal B}_{\overline{B}{}^0\to\rho^-\pi^+}^{\rm eff}
\bigl[s_{+0}^{}\mbox{$>$}s_{-0}^{}\bigr]
\,=\,  16.2\times 10^{-6}   \,\,,  
\hspace{2em}   
{\cal B}_{\overline{B}{}^0\to\rho^+\pi^-}^{\rm eff}
\bigl[s_{+0}^{}\mbox{$<$}s_{-0}^{}\bigr]
\,=\,  4.9\times 10^{-6}   \,\,, 
\end{array}     
\end{eqnarray}     
and 
\begin{eqnarray}   \label{effBb[rhopi]}
\begin{array}{c}   \displaystyle   
{\cal B}_{B^0\to\rho^0\pi^0}^{\rm eff}
\bigl[s_{+0}^{}\mbox{$>$}s_{-0}^{}\bigr]  
\,=\,  0.18\times 10^{-6}   \,\,,  
\hspace{2em}   
{\cal B}_{B^0\to\rho^0\pi^0}^{\rm eff}
\bigl[s_{+0}^{}\mbox{$<$}s_{-0}^{}\bigr]  
\,=\,  0.56\times 10^{-6}   \,\,,  
\vspace{1ex} \\   \displaystyle   
{\cal B}_{\overline{B}{}^0\to\rho^0\pi^0}^{\rm eff}
\bigl[s_{+0}^{}\mbox{$>$}s_{-0}^{}\bigr]  
\,=\,  0.48\times 10^{-6}   \,\,,  
\hspace{2em}   
{\cal B}_{\overline{B}{}^0\to\rho^0\pi^0}^{\rm eff}
\bigl[s_{+0}^{}\mbox{$<$}s_{-0}^{}\bigr]  
\,=\,  0.16\times 10^{-6}   \;.  
\end{array}     
\end{eqnarray}     
For reference, we report 
the two-body branching ratios and the corresponding rate 
asymmetries of the decay modes considered in this paper in Table~\ref{2body}.
We note that the combined branching ratios of the two-body 
$B\to \rho^\pm\pi^\mp$ branching ratios we report are slightly high 
with respect to 
recent measurements~\cite{B->rhopicleo,B->rhopibabar,B->rhopibelle},
but the 
estimates would be consistent with the data if the uncertainties 
in the input parameters were taken into account.  
Returning to our discussion, we 
find that the effective $\rho^0\pi^0$ branching ratios are much 
enhanced with respect to the corresponding two-body rates in 
Table~\ref{2body} --- this is caused by the large interference 
with the  $\rho^\pm\pi^\mp$  contributions.  
The population asymmetries about the $\,s_{+0}^{}=s_{-0}^{}\,$  
line are then   
\begin{eqnarray}  \label{effA[rhopi]}
\begin{array}{c}   \displaystyle   
{\cal A}_{3\pi}^{+-}  \,\equiv\,  
{ \Gamma_{B,\bar B\to\rho^-\pi^+}^{\rm eff}
 \bigl[s_{+0}^{}\mbox{$>$}s_{-0}^{}\bigr]  
 - \Gamma_{B,\bar B\to\rho^+\pi^-}^{\rm eff}
  \bigl[s_{+0}^{}\mbox{$<$}s_{-0}^{}\bigr]  
 \over  
 \Gamma_{B,\bar B\to\rho^-\pi^+}^{\rm eff}
  \bigl[s_{+0}^{}\mbox{$>$}s_{-0}^{}\bigr]
 + \Gamma_{B,\bar B\to\rho^+\pi^-}^{\rm eff}
  \bigl[s_{+0}^{}\mbox{$<$}s_{-0}^{}\bigr] }  
\,=\,  -3.3\%   \,\,,  
\vspace{2ex} \\   \displaystyle   
{\cal A}_{3\pi}^{00}  \,\equiv\,  
{ \Gamma_{B,\bar B\to\rho^0\pi^0}^{\rm eff} 
 \bigl[s_{+0}^{}\mbox{$>$}s_{-0}^{}\bigr]  
 - \Gamma_{B,\bar B\to\rho^0\pi^0}^{\rm eff}
  \bigl[s_{+0}^{}\mbox{$<$}s_{-0}^{}\bigr]    
 \over  
 \Gamma_{B,\bar B\to\rho^0\pi^0}^{\rm eff} 
  \bigl[s_{+0}^{}\mbox{$>$}s_{-0}^{}\bigr]  
 + \Gamma_{B,\bar B\to\rho^0\pi^0}^{\rm eff}
  \bigl[s_{+0}^{}\mbox{$<$}s_{-0}^{}\bigr] }  
\,=\,  -4.6\%   \,\,,     
\end{array}     
\end{eqnarray}     
to be compared with 
${\cal A}_{3\pi}^{(2)}\,\,,{\cal A}_{3\pi}^{(3)}$, respectively, in
Eq.~(\ref{A23pi}).
There is very little  $\Gamma_{3\pi}^{(3)}$  
contribution in  ${\cal A}_{3\pi}^{+-}$, whereas 
in ${\cal A}_{3\pi}^{00}$ 
the 
$\Gamma_{3\pi}^{(2)}$  dominance is much less pronounced, 
yielding  $\,\sim$$-2.6\,$  in the  $\,-4.6.\,$  
Practically, it is challenging to measure  ${\cal A}_{3\pi}^{00}$, 
due to the smallness of the effective  $\rho^0\pi^0$  rates.      
We note that the numerical value of ${\cal A}_{3\pi}^{+-}$ is fairly close to 
that of ${\cal A}_{CP}^{}$ for  $\,B^0\to\rho^+\pi^-\,$  in 
Table~\ref{2body}, and this arises from the dominance of 
the  $\,B^0\to\rho^+\pi^-\,$  and   $\,\bar B^0\to\rho^-\pi^+\,$  
contributions in the $\rho^\pm$ regions of the Dalitz plot.  

\begin{table}[t]   
\caption{\label{2body}%
Estimated two-body branching ratios and rate asymmetries
using the model of Ref.~\cite{AliKL} and the input
parameters given in the Appendix. 
} \centering   \footnotesize
\vskip 0.5\baselineskip
\begin{tabular}{@{\hspace{3ex}}c@{\hspace{5ex}}
c@{\hspace{5ex}}c@{\hspace{5ex}}c@{\hspace{3ex}}}  
\hline \hline  
$\begin{array}{c}\displaystyle \mbox{Decay mode}
 \\ \displaystyle B\to VP \end{array}$  &  
$10^6\,\, {\cal B}_{B\to VP}^{}$  &  
$10^6\,\, {\cal B}_{\bar B\to\bar V\bar P}^{}$  &  
${\cal A}_{CP}^{}={\displaystyle
\Gamma_{\bar B\to\bar V\bar P}^{}-\Gamma_{B\to VP}^{} 
\over\displaystyle    
\Gamma_{\bar B\to\bar V\bar P}^{}+\Gamma_{B\to VP}^{}}\,\,\, 
(\%)  \vphantom{\Biggl]_|^|}$
\\ \hline & && \vspace{-3ex} \\   
$B^0\to\rho^+\pi^-$   &   23.2  &   21.4  &   $-4.1\hphantom{-}$  \\  
$B^0\to\rho^-\pi^+$   &    6.47  &    6.30  &   $-1.3\hphantom{-}$  \\  
$B^0\to\rho^0\pi^0$   &   0.047  &   0.068  &     19  \\  
$B^0\to D^{*+} D^-$   &    376   &    384   &    1.0  \\  
$B^0\to D^{*-} D^+$   &    322   &    323   &    0.3  \\  
$B_s^0\to K^{*+}K^-$  &   3.44   &   4.76   &    16  \\  
$B_s^0\to K^{*-}K^+$  &   0.271  &   0.375  &    16  \\  
$B_s^0\to D_s^{*+}D_s^-$  &  10044  &  10028  &  $-0.08 \hphantom{-}$  \\  
$B_s^0\to D_s^{*-}D_s^+$  &   8628  &   8623  &  $-0.03 \hphantom{-}$  \\   
& && \vspace{-3ex} \\   
\hline \hline  
\end{tabular}
\vspace{1ex} \\  
\end{table}

We expect that  $\,B,\bar B\to \pi^+ \pi^- \pi^0\,$  decay is not populated
exclusively by  $\,B,\bar B\to \rho\pi\,$  contributions; other resonant
and nonresonant contributions ought to occur --- and 
can potentially enter the $\rho\pi$ phase space as well. 
Among the latter, contributions mediated by the  $B^*$ meson 
may be important~\cite{dea&co,DeaPol}. 
Such effects are expected, however, 
to be suppressed by the $BB^*\pi$ vertex, as 
the  $B^*$  is highly off-mass-shell 
in the relevant kinematical region~\cite{gt}, so that we shall
not consider these effects further. 
As for contributions from other resonances, 
we expect the 
``$\sigma$,''  or  $f_0^{}(400-1200)$,  a broad isospin  
$\,I=0\,$  and  spin  $\,J=0\,$  enhancement in  $\pi\pi$ 
scattering,  to play a modest role~\cite{DeaPol,GarMei,gt}.     
The peak of this enhancement is close to the $\rho$  in mass, 
and the  $\sigma\pi^0$  intermediate state contributes 
preferentially to the  $\rho^0\pi^0$  phase space.   
To explore how corrections to $\rho$ ``dominance'' can impact
the population asymmetry, we 
assume that the $\sigma$ resonance is the only 
additional contribution to the  $3\pi$  final states, 
in the  $\rho$  bands on the Dalitz plot.    
 
We express the $\,B\to\sigma\pi^0\,$  amplitudes as  
\begin{eqnarray}   \label{M_pisigma}  
{\cal M}_{B^0\to \pi^0\sigma}^{}  \,=\,  a_\sigma^{}   \,\,,     
\hspace{2em}  
{\cal M}_{\bar B^0\to\pi^0\sigma}^{}  \,=\,  \bar a_\sigma^{}   \;.  
\end{eqnarray}     
In the presence of the $\sigma$, the  $\,B\to3\pi\,$  amplitudes 
in Eq.~(\ref{M_3pi})  
become
\begin{eqnarray}   \label{M_3pi'} 
\begin{array}{c}   \displaystyle   
{\cal M}_{3\pi}^{}  \,=\,   
a_g^{}\, f_g^{} + a_u^{}\, f_u^{} + a_n^{}\, f_n^{}   
+ a_\sigma^{}\, f_\sigma^{}   \,\,,      
\hspace{3em}   
\bar{\cal M}_{3\pi}^{}  \,=\,  
 \bar a_g^{}\, f_g^{} + \bar a_u^{}\, f_u^{} + \bar a_n^{}\, f_n^{}   
+ \bar a_\sigma^{}\, f_\sigma^{}   \,\,,  
\end{array}     
\end{eqnarray}     
where  $f_\sigma^{}$  is the form factor describing the 
$\,\sigma\to\pi^+\pi^-\,$  transition.   
Expressions for the  $\,B\to\sigma\pi\,$  amplitudes   
and  $f_\sigma^{}$  are given in the Appendix.     
The resulting  $\,B\to3\pi\,$  width is
\begin{eqnarray}   \label{Gamma_3pi'}
\Gamma_{3\pi}^{}  &=&  
\int{\rm d}\Phi \left\{  
\sum_{\kappa=g,u,n,\sigma} 
\left( \bigl| a_\kappa^{} \bigr|^2 + \bigl| \bar a_\kappa^{} \bigr|^2 
\vphantom{|_|^|} \right) \bigl| f_\kappa^{} \bigr|^2 
\,+\,  
2\, {\rm Re} \left[ 
\left( a_g^* a_n^{}+\bar a_g^* \bar a_n^{} \right) f_g^* f_n^{} 
+ \left( a_u^* a_\sigma^{}+\bar a_u^* \bar a_\sigma^{} \right) 
 f_u^* f_\sigma^{} 
\vphantom{|_|^|} \right]   
\right\} 
\nonumber \\  && \!\!\!    
+\,\,   
\Gamma_{3\pi}^{(2)} + \Gamma_{3\pi}^{(3)}    
+ \Gamma_{3\pi}^{(4)} + \Gamma_{3\pi}^{(5)}   \,\,,   
\end{eqnarray}     
where  $\Gamma_{3\pi}^{(2,3)}$  are given in 
(\ref{Gamma23})  and   
\begin{eqnarray}   \label{Gamma45}
\begin{array}{c}   \displaystyle   
\Gamma_{3\pi}^{(4)}  \,=\,   
\int{\rm d}\Phi\; 2\, {\rm Re} \left[ 
\left( a_g^* a_\sigma^{}+\bar a_g^* \bar a_\sigma^{} \right) 
f_g^* f_\sigma^{} 
\vphantom{|_|^|} \right]   \,\,,  
\hspace{2em}  
\Gamma_{3\pi}^{(5)}  \,=\,   
\int{\rm d}\Phi\; 2\, {\rm Re} \left[ 
\left( a_n^* a_\sigma^{}+\bar a_n^* \bar a_\sigma^{} \right) 
f_n^* f_\sigma^{} 
\vphantom{|_|^|} \right]   \,\,.   \hspace*{1em}  
\end{array}     
\end{eqnarray}     

With the inclusion of the $\sigma$, we encounter more  
CP-violating observables. Under CP transformations 
we also have\footnote{Noting 
$\,CP|\sigma\rangle = +|\sigma\rangle,\,$  we find 
$\,\langle \pi^0 (p_0) \sigma (p_\sigma)  | Q_i | \bar B^0\rangle 
 = - \eta_B^\ast \eta_Q 
\langle \pi^0 (p_0) \sigma (p_\sigma) | Q_i^\dagger | B^0\rangle \,$
where  $\,p_\sigma = p_+ + p_-\,$  and  $\,\bm{p}_\sigma=\bm{0}.\,$    
With  $\,\eta_B^\ast \eta_Q =+1,\,$  we recover Eq.~(\ref{CP[a_s]}).}   
\begin{eqnarray}   \label{CP[a_s]}
a_\sigma^{}  \,\stackrel{\,CP}{\longleftrightarrow}\,  
-\bar a_\sigma^{}   \,\,,  
\end{eqnarray}     
so that 
$\,a_u^* a_\sigma^{}+\bar a_u^* \bar a_\sigma^{}\,$  
in Eq.~(\ref{Gamma_3pi'}) is CP even,  whereas   
\begin{eqnarray}  \label{ac*as}   
a_g^* a_\sigma^{}+\bar a_g^* \bar a_\sigma^{}  
\hspace{2em} {\rm and} \hspace{2em}   
a_n^* a_\sigma^{}+\bar a_n^* \bar a_\sigma^{}   
\end{eqnarray}     
in Eq.~(\ref{Gamma45}) are CP odd, the latter serving as specific examples 
of the CP-violating amplitude combinations given in Eq.~(\ref{gen:oddl}). 
As we have discussed in conjunction with Eq.~(\ref{ac*ad}), 
a nonzero value of any one of the last two combinations 
signals direct CP violation. 
Noting that  $f_\sigma^{}$  is even under the interchange   
$\,s_{+0}^{}\leftrightarrow s_{-0}^{},\,$  
we can see that both  $f_g^*f_\sigma^{}$  and  $f_n^*f_\sigma^{}$  
are odd under such an interchange.    
It follows that  
\begin{eqnarray}  \label{Gammai'}   
\Gamma_{3\pi}^{(4,5)} \bigl[s_{+0}^{}\mbox{$>$}s_{-0}^{}\bigr] 
\,=\,   
-\Gamma_{3\pi}^{(4,5)} \bigl[s_{+0}^{}\mbox{$<$}s_{-0}^{}\bigr]   \,\,, 
\end{eqnarray}     
so that  $\Gamma_{3\pi}^{(4,5)}$  contribute to the population 
asymmetry on the Dalitz plot about the $\,s_{+0}^{}=s_{-0}^{}\,$  
line,  reflecting direct CP violation, as we have discussed
in Sec.~\ref{dcpv}.
We remark that the CP-odd observables in Eq.~(\ref{ac*as}) share 
with those in Eq.~(\ref{ac*ad}) the unusual property that they 
do not rely on the nonvanishing of the strong phases in $a_i$.   
However, the combinations in Eq.~(\ref{ac*as}) arise not from the 
interference of a pair of CP-conjugate states yielding the ones 
in Eq.~(\ref{ac*ad}), but from the interference of the $\rho^0\pi^0$  
and  $\sigma\pi^0$  intermediate states which have different 
relative angular-momenta,  $\,l=1$  and  $0$,  respectively.  
 
To find the asymmetries in the presence of the  $\sigma$,  
we recompute the effective branching ratios for  
$\,\delta=0.3\,\rm GeV,\,$  as reported in Eqs.~(\ref{effB[rhopi]}) 
and~(\ref{effBb[rhopi]}).      
The results are       
\begin{eqnarray}   \label{effB[rhopi+sigpi]}
\begin{array}{c}   \displaystyle   
{\cal B}_{B^0\to\rho^-\pi^+}^{\rm eff} 
\bigl[s_{+0}^{}\mbox{$>$}s_{-0}^{}\bigr] 
\,=\,  4.8\times 10^{-6}   \,\,,  
\hspace{2em}   
{\cal B}_{B^0\to\rho^+\pi^-}^{\rm eff} 
\bigl[s_{+0}^{}\mbox{$<$}s_{-0}^{}\bigr]  
\,=\,  17.6\times 10^{-6}   \,\,,  
\vspace{1ex} \\   \displaystyle   
{\cal B}_{\overline{B}{}^0\to\rho^-\pi^+}^{\rm eff}  
\bigl[s_{+0}^{}\mbox{$>$}s_{-0}^{}\bigr]
\,=\,  16.2\times 10^{-6}   \,\,,  
\hspace{2em}   
{\cal B}_{\overline{B}{}^0\to\rho^+\pi^-}^{\rm eff}
\bigl[s_{+0}^{}\mbox{$<$}s_{-0}^{}\bigr]
\,=\,  4.8\times 10^{-6}   \,\,, 
\end{array}     
\end{eqnarray}     
\begin{eqnarray}   \label{effBb[rhopi+sigpi]}
\begin{array}{c}   \displaystyle   
{\cal B}_{B^0\to\rho^0\pi^0}^{\rm eff}
\bigl[s_{+0}^{}\mbox{$>$}s_{-0}^{}\bigr]  
\,=\,  0.18\times 10^{-6}   \,\,,  
\hspace{2em}   
{\cal B}_{B^0\to\rho^0\pi^0}^{\rm eff}
\bigl[s_{+0}^{}\mbox{$<$}s_{-0}^{}\bigr]  
\,=\,  0.58\times 10^{-6}   \,\,,  
\vspace{1ex} \\   \displaystyle   
{\cal B}_{\overline{B}{}^0\to\rho^0\pi^0}^{\rm eff}
\bigl[s_{+0}^{}\mbox{$>$}s_{-0}^{}\bigr]  
\,=\,  0.51\times 10^{-6}   \,\,,  
\hspace{2em}   
{\cal B}_{\overline{B}{}^0\to\rho^0\pi^0}^{\rm eff}
\bigl[s_{+0}^{}\mbox{$<$}s_{-0}^{}\bigr]  
\,=\,  0.16\times 10^{-6}   \,\,,  
\end{array}     
\end{eqnarray}     
which indicate that the effects of the $\sigma$  
in the $\rho^\pm \pi^\mp$ bands 
are really very small. 
The population asymmetries about the $\,s_{+0}^{}=s_{-0}^{}\,$  
line are then   
\begin{eqnarray}   \label{effA[rhopi+sigpi]}
\begin{array}{c}   \displaystyle   
{\cal A}_{3\pi}^{+-}  \,=\,  
{ \Gamma_{B,\bar B\to\rho^-\pi^+}^{\rm eff}
 \bigl[s_{+0}^{}\mbox{$>$}s_{-0}^{}\bigr]  
 - \Gamma_{B,\bar B\to\rho^+\pi^-}^{\rm eff}
  \bigl[s_{+0}^{}\mbox{$<$}s_{-0}^{}\bigr]  
 \over  
 \Gamma_{B,\bar B\to\rho^-\pi^+}^{\rm eff}
  \bigl[s_{+0}^{}\mbox{$>$}s_{-0}^{}\bigr]
 + \Gamma_{B,\bar B\to\rho^+\pi^-}^{\rm eff}
  \bigl[s_{+0}^{}\mbox{$<$}s_{-0}^{}\bigr] }  
\,=\,  -3.3\%   \,\,,  
\vspace{2ex} \\   \displaystyle   
{\cal A}_{3\pi}^{00}  \,=\,  
{ \Gamma_{B,\bar B\to\rho^0\pi^0}^{\rm eff} 
 \bigl[s_{+0}^{}\mbox{$>$}s_{-0}^{}\bigr]  
 - \Gamma_{B,\bar B\to\rho^0\pi^0}^{\rm eff}
  \bigl[s_{+0}^{}\mbox{$<$}s_{-0}^{}\bigr]    
 \over  
 \Gamma_{B,\bar B\to\rho^0\pi^0}^{\rm eff} 
  \bigl[s_{+0}^{}\mbox{$>$}s_{-0}^{}\bigr]  
 + \Gamma_{B,\bar B\to\rho^0\pi^0}^{\rm eff}
  \bigl[s_{+0}^{}\mbox{$<$}s_{-0}^{}\bigr] }  
\,=\,  -3.0\%   \;.    
\end{array}     
\end{eqnarray}     
Comparing the  ${\cal A}_{3\pi}^{+-}$  values in 
Eqs.~(\ref{effA[rhopi]})  and~(\ref{effA[rhopi+sigpi]})  
with  ${\cal A}_{3\pi}^{}$  in  Eq.~(\ref{A_3pi}), 
we see that  ${\cal A}_{3\pi}^{+-}$  captures 
the CP-violating effects occurring in the dominant 
$\rho^\pm\pi^\mp$  modes, as encoded in $\Gamma_{3\pi}^{(2)}$.   
In contrast, the  ${\cal A}_{3\pi}^{00}$  value in 
Eq.~(\ref{effA[rhopi+sigpi]}) receives the respective contributions 
$\,(-2.6-2.0+1.9-0.3)\%\,$  from  $\Gamma_{3\pi}^{(2+3+4+5)}$, 
so that the presence of the $\sigma$ leads to a sizable reduction of 
the asymmetry. The interpretation of this latter asymmetry is thus
rendered more difficult. 

The untagged asymmetry, in specific, ${\cal A}_{3\pi}^{+-}$, has
been investigated by BaBar and Belle. BaBar reports~\cite{B->rhopibabar}
\begin{equation}
{\cal A}_{CP}^{\rho\pi}
= \frac{\Gamma_{B,\bar B\to\rho^+\pi^-}^{\rm eff} - 
\Gamma_{B,\bar B\to\rho^-\pi^+}^{\rm eff}}
{ \Gamma_{B,\bar B\to\rho^-\pi^+}^{\rm eff}
 + \Gamma_{B,\bar B\to\rho^+\pi^-}^{\rm eff}}
=-0.18 \pm 0.08 \pm 0.03 \,\,,
\end{equation}
with the recent update 
${\cal A}_{CP}^{\rho\pi}=-0.11 \pm 0.06 \pm 0.03 \,$~\cite{rhopibabar2}.
The $s_{+0} {}^{>}_{<} s_{-0} $ criterion we impose in
Eq.~(\ref{effA[rhopi]}) has apparently not been effected
in ${\cal A}_{CP}^{\rho\pi}$. Thus this quantity may treat
events in the region where the $\rho^\pm$ bands overlap inappropriately. 
In the theoretical analysis we report in Table~\ref{G23}, 
the impact of the overlap region, which contains an intrinsically
larger asymmetry, is diluted by its small contribution
to the total $B\to \rho^\pm\pi^\mp$ decay rate. Thus
the $s_{+0} {}^{>}_{<} s_{-0} $ correction may not be essential; 
${\cal A}_{CP}^{\rho\pi}$ may well be tantamount to $-{\cal A}_{3\pi}^{+-}$. 
The Belle measurement, in contrast, removes
events with an ambiguous $\rho$ charge assignment 
from their data set~\cite{B->rhopibelle}; this is a practical 
realization of our $s_{+0} {}^{>}_{<} s_{-0} $ criterion.
Belle reports~\cite{B->rhopibelle}
\begin{equation}
{\cal A} = -0.38^{+0.19}_{-0.21}\hbox{(stat)}^{+0.04}_{-0.05}\hbox{(syst)}\,\,,
\end{equation}
where  $\,{\cal A} = - {\cal A}_{3\pi}^{+-}$.

The central values of the BaBar and Belle measurements, albeit
their still-sizeable errors, point to much larger 
CP-violating effects than what we have calculated. Interestingly,
more sophisticated calculations show similar trends. 
For example, 
${\cal A}_{3\pi}^{+-}$ 
has been calculated in the QCD factorization
approach to yield the ``default results''~\cite{BenNeu}
\begin{equation}
{\cal A}_{3\pi}^{+-} = 
-0.01^{+0.00 + 0.01 + 0.00 + 0.10}_{-0.00 - 0.01 -0.00 -0.10}\,\,,
\end{equation}
where larger, more positive asymmetries, outside the stated error
ranges, can occur
in various excursions from their default assumptions~\cite{BenNeu}.
Here there is no 
subtlety concerning the ``overlap regions'' as their 
assay assumes the $\rho$ meson has zero-width. Using the results
of Table~\ref{2body}, we find in comparison that 
$\,{\cal A}_{3\pi}^{+-} = -0.028\,$  in this limit, so that our numerical
results are of comparable size. 
Moreover, the size of the partial-rate
asymmetry in the  $\,B\to \rho^+\pi^-\,$  mode is 
crudely commensurate with the size of ${\cal A}_{3\pi}^{+-}$~\cite{BenNeu}, 
as we have found in 
our calculation, so that one would expect that a search for DCPV in
these modes through the population asymmetry would be statistically
more efficacious.

This concludes our discussion of  $\,B\to\pi^+\pi^-\pi^0\,$  decay. 
In what follows we consider how the study of the population asymmetry
in other modes can complement tests of the SM and searches for
emergent, new phenomena.
For example, the time-dependent 
asymmetry associated with the decay of a state tagged as
$B^0$ or $\bar B^0$ into a CP eigenstate $f$ is characterized by
two parameters:  $C_f^{}$  and  $S_f^{}$. The first term is generated 
through direct CP violation, whereas the second is 
generated through the interference of $B$-$\bar B$ mixing
and direct decay. In the SM,  $S_f^{}$ 
measures $-\eta_f\sin(2\beta)$, where  $\,\eta=+1\, (-1)\,$ for a CP-even (odd)
final state\footnote{We define the asymmetry as 
$A(t) \equiv (\Gamma(\bar B^0\to f) - \Gamma(B^0\to f))/(\Gamma(\bar B^0\to f) 
+ \Gamma(B^0\to f)) \equiv - C_f \cos(\Delta M\, t) + S_f \sin(\Delta M\, t).$
}, in modes driven by quark-level transitions such as 
$b\to c\bar c s\,$, $b\to s \bar s s\,$, and $b\to c\bar c d\,$, 
modulo small corrections~\cite{GroWor}. 
The determination of $\sin(2\beta)$ from  $S_{J/\psi K_S}$ serves
as a benchmark against which $S_f^{}$, from other modes, can be measured. 
The modes  $\,B\to \phi K_S^{},\,$  $\,B\to K_S^{} K^+ K^-\,$  decays,  
mediated via  $\,b\to s\bar s s,\,$  $\,B\to D^{*+}D^-\,$  decay,  
mediated via  $\,b\to c\bar c d,\,$  and  $\,B_s^{}\to D_s^{*+}D_s^-\,$  decay, 
mediated via  $\,b\to c\bar c s,\,$  serve as
examples. Of these, only $\phi K_S^{}$ is a CP-eigenstate. One can,
nevertheless, determine $S_f^{}$ from the three-body decays, 
of which the vector-pseudoscalar modes are part; 
in these cases the coefficient 
of the $\sin(\Delta M\,t)$ term must be corrected for a 
non-CP-violating dilution factor to yield 
$S_f^{}$~\cite{gronau89,aleksan91,aleksan93,babarbook}. 
We focus on the three-body decays as the population asymmetry
therein allows us to test the extent to which the corrections
to the SM tests suggested in Ref.~\cite{GroWor} really are small.
Alternatively, such observables may well help us establish 
the character of the new physics which induce deviations from the SM 
predictions. In the following we consider 
$\,B\to D^{*+}D^- \to D^+D^-\pi^0\,$  and  
$\,B_s^{}\to D_s^{*+}D_s^- \to D_s^+ D_s^- \pi^0\,$ 
decays as an explicit realization of these ideas; we also include 
$\,B_s^{}\to K^{*+}K^- \to K^+ K^- \pi^0\,$  decays as well to show
that the population asymmetry can also be quite large in the SM.

Before turning to these examples, we note that the population asymmetry
may well serve as a useful tool in the study of  
$\,B\to K_S^{} K^+ K^-\,$  decay.
At issue is the extent to which the  $K_S^{} K^+ K^-$ final state is 
CP-even once the regions associated with the $\phi$ resonance are 
explicitly removed from consideration~\cite{B->KKK,Grossman:2003qp,Gronau:bu}. 
We note as per Refs.~\cite{Grossman:2003qp,Gronau:bu} that the 
contributions of CP-even and CP-odd final states 
to  $\,B\to\bigl[K^+ K^-\bigr]_l \bigl(K_S^{}\bigr)_l\,$  
decay differ in their angular-momentum character; their interference can 
give rise to a direct CP-violating population asymmetry as described in 
Eq.~(\ref{gen:oddl}). The population asymmetry serves as 
an independent empirical assay of the reliability of the underlying
assumptions of their analysis~\cite{B->KKK,Grossman:2003qp,Gronau:bu}.

\subsection{\mbox{\boldmath$B\to D^+D^-\pi^0\,$}\label{DDpi}}

We write the amplitudes for  
$\,B^0,\bar B^0\to D^{*+}D^-,D^{*-}D^+\,$  as   
\begin{eqnarray}   \label{M_D*D} 
\begin{array}{c}   \displaystyle   
{\cal M}_{B^0\to D^{*k}D^l}^{}  \,=\,  
a_{kl}^{}\, \varepsilon_{D^*}^*\cdot p_D^{}   \,\,,     
\hspace{2em}  
{\cal M}_{\bar B^0\to D^{*k}D^l}^{}  \,=\,     
\bar a_{kl}^{}\, \varepsilon_{D^*}^*\cdot p_D^{}   \,\,,  
\end{array}     
\end{eqnarray}     
where the expressions for $a_{kl}^{}$  and  $\bar a_{kl}^{}$  
are given in the Appendix.   
The amplitudes for 
$\,B^0,\bar B^0\to D^+ \bigl( p_+^{} \bigr) \,  
D^- \bigl( p_-^{} \bigr) \, \pi^0 \bigl( p_0^{} \bigr) \,$    
are then   
\begin{eqnarray}   \label{M_DDpi} 
\begin{array}{c}   \displaystyle   
{\cal M}_{D\bar D\pi}^{}  \,=\,  a_{+-}^{}\, f_+^{} + a_{-+}^{}\, f_-^{} 
\,=\,  a_g^{}\, f_g^{} + a_u^{}\, f_u^{}   \,\,,      
\vspace{2ex} \\   \displaystyle   
\bar {\cal M}_{D\bar D\pi}^{}  \,=\,  
\bar a_{+-}^{}\, f_+^{} + \bar a_{-+}^{}\, f_-^{} 
\,=\,  \bar a_g^{}\, f_g^{} + \bar a_u^{}\, f_u^{}   \;.    
\end{array}     
\end{eqnarray}     
Here, as in the  $3\pi$  case, 
\begin{eqnarray}   \label{aba'}
\begin{array}{c}   \displaystyle   
a_g^{}  \,=\,  a_{+-}^{}+a_{-+}^{}   \,\,,   
\hspace{2em}  
a_u^{}  \,=\,  a_{+-}^{}-a_{-+}^{}   \,\,,   
\hspace{2em}  
\bar a_g^{}  \,=\,  \bar a_{+-}^{}+\bar a_{-+}^{}   \,\,,   
\hspace{2em}  
\bar a_u^{}  \,=\,  \bar a_{+-}^{}-\bar a_{-+}^{}   \,\,,   
\vspace{1ex} \\   \displaystyle   
2\, f_g^{}  \,=\,  f_+^{}+f_-^{}   \,\,,   
\hspace{2em}  
2\, f_u^{}  \,=\,  f_+^{}-f_-^{}   \,\,,   
\end{array}     
\end{eqnarray}     
in this case, however, the $f_\pm^{}$  are given by Eq.~(\ref{f+-}) in the 
Appendix.   
The  $\,B^0,\bar B^0\to D^+ D^-\pi^0\,$  amplitudes lead to 
the decay width 
\begin{eqnarray}   
\Gamma_{D\bar D\pi}^{}  \,=\,  
\int {\rm d}\Phi 
\left( \left| \vphantom{\bar M}{\cal M}_{D\bar D\pi}^{} \right|^2 
      + \left| \bar{\cal M}_{D\bar D\pi}^{} \right|^2 \right) 
\,=\,  \Gamma_{D\bar D\pi}^{(1)} + \Gamma_{D\bar D\pi}^{(2)}   \,\,,  
\end{eqnarray}     
where   
\begin{eqnarray}  \label{Gamma23_DDp}   
\begin{array}{c}   \displaystyle   
\Gamma_{D\bar D\pi}^{(1)}  \,=\,   
\int {\rm d}\Phi \sum_{\kappa=g,u} 
\left( \left| a_\kappa^{} \right|^2 + \left| \bar a_\kappa^{} \right|^2 
\vphantom{|_|^|} \right) \bigl| f_\kappa^{} \bigr|^2   \,\,,   
\hspace{2em}    
\Gamma_{D\bar D\pi}^{(2)}  \,=\,   
\int {\rm d}\Phi\; 2\, {\rm Re} \left[ 
\left( a_g^* a_u^{}+\bar a_g^* \bar a_u^{} \right) f_g^* f_u^{} 
\vphantom{|_|^|} \right]   \,\,.  
\hspace*{1em}    
\end{array}     
\end{eqnarray}     
It follows that  
\begin{eqnarray}  \label{Gammai_DDp}   
\Gamma_{D\bar D\pi}^{(1)} \bigl[s_{+0}^{}\mbox{$>$}s_{-0}^{}\bigr] 
\,=\,   
+\Gamma_{D\bar D\pi}^{(1)} \bigl[s_{+0}^{}\mbox{$<$}s_{-0}^{}\bigr]   \,\,,  
\hspace{2em}  
\Gamma_{D\bar D\pi}^{(2)} \bigl[s_{+0}^{}\mbox{$>$}s_{-0}^{}\bigr] 
\,=\,   
-\Gamma_{D\bar D\pi}^{(i)} \bigl[s_{+0}^{}\mbox{$<$}s_{-0}^{}\bigr]   \,\,.  
\end{eqnarray}     
The CP-odd combination  $\,a_g^* a_u^{}+\bar a_g^*\bar a_u^{}\,$  
in this case is proportional to  $\,\sin\beta,\,$  as can be 
inferred from the CKM factors in the $\,B\to D^{*\pm}D^\mp\,$  
amplitudes given in the Appendix.  
    
To evaluate  $\Gamma_{D\bar D\pi}^{(1,2)}$,  we consider 
the Dalitz plot of $s_{+0}^{}$  vs.  $s_{-0}^{}$  for 
the  $D^+D^-\pi^0$  final states.  
Integrating over the two regions divided by 
the $\,s_{+0}^{}=s_{-0}^{}\,$  line, we find    
\begin{eqnarray}   \label{DDp} 
\begin{array}{c}   \displaystyle   
{\cal B}_{D\bar D\pi}^{(2)}\bigl[s_{+0}^{}\mbox{$>$}s_{-0}^{}\bigr]  
\,=\,  1.0\times 10^{-6}   \,\,,  
\hspace{3em}  
{\cal B}_{D\bar D\pi}^{}  \,=\,  
2\, {\cal B}_{D\bar D\pi}^{(1)}\bigl[s_{+0}^{}\mbox{$>$}s_{-0}^{}\bigr]  
\,=\,  4.8\times 10^{-4}   \,\,,    
\end{array}     
\end{eqnarray}     
where  
$\,{\cal B}_{D\bar D\pi}^{(i)}\equiv
\Gamma_{D\bar D\pi}^{(i)}/\Gamma_{B^0}^{}\,$  
for  $\,i=1,2.\,$  
The resulting CP-violating asymmetry is  
\begin{eqnarray}   \label{A_DDpi} 
\begin{array}{c}   \displaystyle   
{\cal A}_{D\bar D\pi}^{}  \,=\,   
{ \Gamma_{D\bar D\pi}^{}\bigl[s_{+0}^{}\mbox{$>$}s_{-0}^{}\bigr]
 - \Gamma_{D\bar D\pi}^{}\bigl[s_{+0}^{}\mbox{$<$}s_{-0}^{}\bigr] 
 \over  
 \Gamma_{D\bar D\pi}^{}\bigl[s_{+0}^{}\mbox{$>$}s_{-0}^{}\bigr]
 + \Gamma_{D\bar D\pi}^{}\bigl[s_{+0}^{}\mbox{$<$}s_{-0}^{}\bigr] }
\,=\,   
{ \Gamma_{D\bar D\pi}^{(2)}\bigl[s_{+0}^{}\mbox{$>$}s_{-0}^{}\bigr]
 \over    
 \Gamma_{D\bar D\pi}^{(1)}\bigl[s_{+0}^{}\mbox{$>$}s_{-0}^{}\bigr] }
\,=\,  +0.4\%   \;.  
\end{array}     
\end{eqnarray}     
More realistically, the  $D^+D^-\pi^0$  final states also 
receive contributions from heavier charmed resonances such as 
the  $D_0^{}$ and  $D_2^{}$  mesons~\cite{D_02}.  
We can isolate the $D^*$ effects, however, by limiting 
the integration regions to the  $D^{*\pm}$  bands.  
Therefore, we calculate the effective rates   
\begin{eqnarray}   \label{Geff_DDpi}
\begin{array}{c}   \displaystyle   
\Gamma_{B\to D^*(p_1+p_2) \bar D(p_3)}^{\rm eff}    
\,=\,  
\Gamma\bigl(B\to \pi^0(p_1) D(p_2)\bar D(p_3)\bigr) \Bigl|_{
(M_{D^*}^{}-\delta)^2\le s_{12}^{}\le (M_{D^*}^{}+\delta)^2}^{}  \,\,,  
\end{array}     
\end{eqnarray}     
in analogy to the effective  $\,B\to3\pi\,$  rates in 
Eq.~(\ref{Geff}).   
Since the $D^*$ width is very narrow, we can choose  
$\,\delta=5\,{\rm MeV}\simeq 50\,\Gamma_{D^{*+}}^{},\,$   
which allows us to capture the $D^*$ contributions 
and avoid those of $D_{0,2}^{}$, which are about  $400\,\rm MeV$  
heavier than  $D^*$.    Thus we obtain  
\begin{eqnarray}   \label{effB[D*D]}  
\begin{array}{c}   \displaystyle   
{\cal B}_{B^0\to D^{*-} D^+}^{\rm eff}  
\bigl[s_{+0}^{}\mbox{$>$}s_{-0}^{}\bigr]  
\,=\,  0.98\times 10^{-4}   \,\,,  
\hspace{2em}   
{\cal B}_{B^0\to D^{*+} D^-}^{\rm eff}  
\bigl[s_{+0}^{}\mbox{$<$}s_{-0}^{}\bigr]  
\,=\,  1.14\times 10^{-4}   \,\,,  
\vspace{2ex} \\   \displaystyle   
{\cal B}_{\overline{B}{}^0\to D^{*-} D^+}^{\rm eff}   
\bigl[s_{+0}^{}\mbox{$>$}s_{-0}^{}\bigr]   
\,=\,  1.16\times 10^{-4}   \,\,,  
\hspace{2em}   
{\cal B}_{\overline{B}{}^0\to D^{*+} D^-}^{\rm eff}  
\bigl[s_{+0}^{}\mbox{$<$}s_{-0}^{}\bigr]  
\,=\,  0.98\times 10^{-4}   \,\,,
\end{array}     
\end{eqnarray}     
where we note that our effective  $D\bar D\pi$  rates are 
about 3 times smaller than the corresponding  two-body  rates given
in Table~\ref{2body}, reflecting the empirical fact that 
$\,\Gamma_{D^{*+}\to D\pi}^{}\simeq 3\,\Gamma_{D^{*+}\to D^+\pi^0}^{}.\,$    

The population asymmetry about the $\,s_{+0}^{}=s_{-0}^{}\,$  line 
becomes 
\begin{eqnarray}   \label{effA[D*D]}
\begin{array}{c}   \displaystyle   
{\cal A}_{D\bar D\pi}^{+-}  \,=\,   
{ \Gamma_{B,\bar B\to D^{*-} D^+}^{\rm eff}  
 \bigl[s_{+0}^{}\mbox{$>$}s_{-0}^{}\bigr] 
 - \Gamma_{B,\bar B\to D^{*+} D^-}^{\rm eff}  
  \bigl[s_{+0}^{}\mbox{$<$}s_{-0}^{}\bigr] 
 \over  
 \Gamma_{B,\bar B\to D^{*-} D^+}^{\rm eff} 
 \bigl[s_{+0}^{}\mbox{$>$}s_{-0}^{}\bigr] 
 + \Gamma_{B,\bar B\to D^{*+} D^-}^{\rm eff} 
  \bigl[s_{+0}^{}\mbox{$<$}s_{-0}^{}\bigr] }
\,=\,  +0.4\%   \,\,,  
\end{array}     
\end{eqnarray}     
which is the same as the result in Eq.~(\ref{A_DDpi}).  
Since the $D^*$ is so narrow, the population asymmetry can also be 
calculated directly from the 
the two-body  rates in Table~\ref{2body}, 
namely,  
\begin{eqnarray}   \label{A[D*D]}
\begin{array}{c}   \displaystyle   
{\cal A}_{D\bar D\pi}^{+-}  \,\simeq\,   
{ \Gamma_{B^0\to D^{*-}D^+}^{}+\Gamma_{\overline{B}{}^0\to D^{*-}D^+}^{}  
 - \Gamma_{B^0\to D^{*+}D^-}^{}-\Gamma_{\overline{B}{}^0\to D^{*+}D^-}^{}  
 \over  
 \Gamma_{B^0\to D^{*-}D^+}^{}+\Gamma_{\overline{B}{}^0\to D^{*-}D^+}^{}  
 + \Gamma_{B^0\to D^{*+}D^-}^{}+\Gamma_{\overline{B}{}^0\to D^{*+}D^-}^{} }   
\,=\,  +0.4\%   \,\,.
\end{array}     
\end{eqnarray}     
The  $D^{*\pm}D^\mp$  rates in this table 
are roughly consistent with data~\cite{B->D*D,Aubert:2003ca}; 
in particular, we note the most precise measurement:
${\cal B}(B\to D^{*\pm} D^\mp)= (8.8 \pm 1.0 \pm 1.3)
\cdot 10^{-4}$~\cite{Aubert:2003ca}. In this reference the
time-integrated asymmetry is reported as well~\cite{Aubert:2003ca}:
\begin{equation}    \label{acpD*D}
{\cal A}=\frac{N_{D^{*+}D^-} - N_{D^{*-}D^+}}{N_{D^{*+}D^-} 
+ N_{D^{*-}D^+}}=-0.03 \pm 0.11 \pm 0.05 \,\,.
\end{equation}
The Dalitz plot associated with the decay 
pathway $B\to D^{*\pm} D^{\mp}\to D^+D^-\pi^0$ gives rise to a
CP-violating population asymmetry, though 
$\,\Gamma_{D^{*+}\to D^0\pi^{+}}\simeq 2\,\Gamma_{D^{*+}\to D^+\pi^0}^{}.\,$ 
In this case, however, as we have seen, the $D^*$ is so narrow that 
the population asymmetry is numerically indistinguishable 
from the untagged asymmetry computed from the two-body decay rates
--- there is no overlap region to treat.
Consequently,
the result of Eq.~(\ref{acpD*D}) can be compared with our estimate
of the population asymmetry, to which it agrees, within errors. 
$S_{D^*D}$ should yield $\sin(2\beta)$ to a good 
approximation~\cite{GroWor}.

\subsection{\mbox{\boldmath$B_s^{}\to D_s^+D_s^-\pi^0\,$}\label{DsDspi}}
    
We now consider the decays  
$\,B_s^0,\bar B_s^0\to D_s^{*\pm}D_s^\mp\to D_s^+D_s^-\pi^0,\,$ in
furtherance of the SM tests of Ref.~\cite{GroWor}. 
As well known, 
the pertinent penguin contribution in 
$\,B_s^0,\bar B_s^0\to D_s^{*\pm}D_s^\mp$ decay 
possesses a subdominant weak phase,  
so that we expect its impact to be very small --- $S_{D_s^*D_s^{}}$ 
should yield 
$\sin(2\beta)$ to an excellent approximation. 
Were  $S_{D_s^*D_s^{}}$  to differ significantly from $\sin(2\beta)$, the
associated direct CP-violating observables would yield helpful
insight into the nature of the emergent phenomena. 
The requisite formulas can be obtained from those in the    
$\,B\to D^{*\pm} D^\mp\to D^\pm D^\mp\pi^0\,$  case by making 
the replacements
\begin{eqnarray}    
d   \,\to\,  s   \,\,,   \hspace{2em}  
B  \,\to\,  B_s^{}   \,\,,   \hspace{2em}  
D^{(*)}  \,\to\,  D_s^{(*)}   \,\,   
\end{eqnarray}     
and by using the  $\,b\to s\,$  and  $\bar b\to\bar s$  entries 
in Table~\ref{a_i}, so that we do not report them explicitly. 
The CKM factors in the  $\,B_s^0\to D_s^{*\pm}D_s^\mp\,$  
amplitudes are  $\,V_{cb}^*V_{cs}^{}\simeq A\lambda^2\,$  and  
$\,V_{tb}^*V_{ts}^{}\simeq-A\lambda^2,\,$
which are largely real, resulting in small CP violation.   
The  $D_s^*$  width has not been measured, but it is expected 
to be extremely small~\cite{GoiRob}, of the order of $0.2\,\rm keV.\,$  
As a consequence, the population asymmetry about the  
$\,s_{+0}^{}=s_{-0}^{}\,$  line on the Dalitz plot for the  
$D_s^+D_s^-\pi^0$  final states can be calculated by using 
the  $B_s^{}\to D_s^*D_s^{}\,$  rates listed in Table~\ref{2body}.  
Thus, 
\begin{eqnarray}   \label{effA[Ds*Ds]}
{\cal A}_{D_s^{}\bar D_s^{}\pi}^{+-}  \,\simeq\,   
{ \Gamma_{B_s^{}\to D^{*-}D^+}^{} 
 + \Gamma_{\overline{B}{}_s^{}\to D^{*-}D^+}^{} 
 - \Gamma_{B_s^{}\to D^{*+}D^-}^{} 
 - \Gamma_{\overline{B}{}_s^{}\to D^{*+}D^-}^{} 
 \over  \Gamma_{B_s^{}\to D^{*-}D^+}^{} 
 + \Gamma_{\overline{B}{}_s^{}\to D^{*-}D^+}^{} 
 + \Gamma_{B_s^{}\to D^{*+}D^-}^{} 
 + \Gamma_{\overline{B}{}_s^{}\to D^{*+}D^-}^{} }  
\,=\,  -0.03\%   \,\,,    
\end{eqnarray}     
which is very small, in accord with SM expectations.  
   
\subsection{\mbox{\boldmath$B_s^{}\to K^+K^-\pi^0\,$}\label{KKpi}}
    
As a final example, we consider 
$\,B_s^0,\bar B_s^0\to K^{*\pm}K^\mp\to K^+K^-\pi^0;\,$
the pertinent formulas are completely analogous to 
those in the  $\,B\to D^+D^-\pi^0\,$  case.  
We need only make the replacements  
$\,c\to u\,$  and  $\,d\to s\,$  for the quarks, 
and  $\,B\to B_s^{}\,$  and  $\,D\to K\,$  
for the mesons.   
The CP-odd combination  
$\,a_g^* a_u^{}+\bar a_g^* \bar a_u^{}\,$  in this case 
is proportional to  $\,\sin\gamma,\,$  as can be inferred from 
the $\,B_s\to K^{*\pm}K^\mp\,$  amplitudes in the Appendix.  
  
Integrating over the two regions on the Dalitz plot 
for the  $\,K^+K^-\pi^0\,$  final states,  
assuming no other contributions,   we find    
\begin{eqnarray}   \label{Gamma{23}K}
\begin{array}{c}   \displaystyle   
{\cal B}_{K\bar K\pi}^{(2)}\bigl[s_{+0}^{}\mbox{$>$}s_{-0}^{}\bigr]  
\,=\,  2.1\times 10^{-7}   \,\,,  
\hspace{2em}  
{\cal B}_{K\bar K\pi}^{}  \,=\,    
2\, {\cal B}_{K\bar K\pi}^{(1)}\bigl[s_{+0}^{}\mbox{$>$}s_{-0}^{}\bigr]  
\,=\,  2.9\times 10^{-6}   \,\,,    
\end{array}     
\end{eqnarray}     
which leads to 
\begin{eqnarray}   \label{A_KKpi}
{\cal A}_{K\bar K\pi}^{}  \,=\,   
{ \Gamma_{K\bar K\pi}^{(2)}\bigl[s_{+0}^{}\mbox{$>$}s_{-0}^{}\bigr]  
 \over    
 \Gamma_{K\bar K\pi}^{(1)}\bigl[s_{+0}^{}\mbox{$>$}s_{-0}^{}\bigr] }  
\,=\,  +15\%   \;.  
\end{eqnarray}     
As we have discussed, it is better to consider 
effective branching ratios instead.   
Thus, taking  $\,\delta=3\Gamma_{K^*}^{},\,$  we obtain   
\begin{eqnarray}   \label{effB[K*K]}  
\begin{array}{c}   \displaystyle   
{\cal B}_{B_s^0\to K^{*-}K^+}^{\rm eff}  
\bigl[s_{+0}^{}\mbox{$>$}s_{-0}^{}\bigr]  
\,=\,  0.09\times 10^{-6}   \,\,,  
\hspace{2em}   
{\cal B}_{B_s^0\to K^{*+}K^-}^{\rm eff}  
\bigl[s_{+0}^{}\mbox{$<$}s_{-0}^{}\bigr]  
\,=\,  1.00\times 10^{-6}   \,\,,  
\vspace{2ex} \\   \displaystyle   
{\cal B}_{\overline{B}{}_s^0\to K^{*-}K^+}^{\rm eff}
\bigl[s_{+0}^{}\mbox{$>$}s_{-0}^{}\bigr]  
\,=\,  1.40\times 10^{-6}   \,\,,  
\hspace{2em}   
{\cal B}_{\overline{B}{}_s^0\to K^{*+}K^-}^{\rm eff} 
\bigl[s_{+0}^{}\mbox{$<$}s_{-0}^{}\bigr]   
\,=\,  0.10\times 10^{-6}   \,\,,  
\end{array}     
\end{eqnarray}     
where the effective  $K\bar K\pi$  rates are 
roughly 3 times smaller than the corresponding  two-body  rates 
in Table~\ref{2body},  reflecting the isospin relation  
$\,\Gamma_{K^{*+}\to K\pi}^{}=3\,\Gamma_{K^{*+}\to K^+\pi^0}^{}.\,$    
We thus find the population asymmetry 
\begin{eqnarray}   \label{effA[K*K]}
\begin{array}{c}   \displaystyle   
{\cal A}_{K\bar K\pi}^{+-}  \,=\,   
{ \Gamma_{B_s^{},\bar B_s^{}\to K^{*-}K^+}^{\rm eff}
 \bigl[s_{+0}^{}\mbox{$>$}s_{-0}^{}\bigr]
 - \Gamma_{B_s^{},\bar B_s^{}\to K^{*+}K^-}^{\rm eff}
   \bigl[s_{+0}^{}\mbox{$<$}s_{-0}^{}\bigr]
 \over  
 \Gamma_{B_s^{},\bar B_s^{}\to K^{*-}K^+}^{\rm eff}
 \bigl[s_{+0}^{}\mbox{$>$}s_{-0}^{}\bigr]
 + \Gamma_{B_s^{},\bar B_s^{}\to K^{*+}K^-}^{\rm eff}
   \bigl[s_{+0}^{}\mbox{$<$}s_{-0}^{}\bigr] }   
\,=\,  +15\%   \;.
\end{array}     
\end{eqnarray}     
This shows, once again, that focusing on the population asymmetry 
in the resonance regions on the Dalitz plot is sufficient 
to capture the CP-violating effect of interest.
We note that our ${\cal A}_{K\bar K\pi}^{+-}$ result is 
close to the value of ${\cal A}_{CP}^{}$ for  
$\,B_s^0\to K^{*+} K^-\,$ decay in Table~\ref{2body}, and this is 
caused by the dominance of the  $\,B_s^0\to K^{*+}K^-\,$  
and  $\,\bar B_s^0\to K^{*-}K^+\,$  contributions to the $K^{*\pm}$ 
bands in the Dalitz plot.

\section{Conclusions\label{conclusion}}    

The study of direct CP violation in the 
$B$-meson system is needed to clarify the nature and origin of CP
violation in nature. In this paper, we have studied a new tool:
the population asymmetry in the untagged $B_d,B_s$ decay 
rate to a self-conjugate final state of three pseudoscalar mesons. 
The population asymmetry emerges from the failure 
of mirror symmetry across the Dalitz plot; it signals
the presence of direct CP violation. 
The amplitudes which give rise to such an asymmetry are of two,
distinct classes. The first, proposed in Ref.~\cite{gardner}, 
follows if we can separate the self-conjugate final state, via
the resonances which appear, into distinct, CP-conjugate states; 
we term the latter CP-enantiomers.  
We note that such states are 
distinguishable by their location with respect to the mirror line of
the Dalitz plot; in this sense they are distinguishable, mirror
images. In this construction the angular momentum of the resonance with
respect to the bachelor meson is fixed; such amplitudes
form part of a more
general class of amplitudes, for which the angular momentum of the
two interfering amplitudes are of the same parity, 
as delineated in Eq.~(\ref{gen:evenl}). 
The second class
of amplitudes emerges from interfering amplitudes whose angular momentum
is of differing parity, as delineated in Eq.~(\ref{gen:oddl}). 
As an explicit example of the latter, we studied 
the CP-violation consequent
to $s$-wave and $p$-wave interference, 
mediated by the $\rho$ and $\sigma$ resonances,
in  $\,B\to\pi^+\pi^-\pi^0\,$  decay. In our numerical examples, 
we find that the population asymmetry in the untagged decay rate 
is typically comparable in size
to the partial rate asymmetries, 
making its measurement statistically
advantageous.

The population asymmetry serves as a complementary assay to the
study of the time-dependent asymmetry in neutral $B$-meson decays. 
In this regard, we have emphasized the SM tests discussed
in Ref.~\cite{GroWor}, specifically the equality of the
quantity $S_f^{}$, induced by the interference of $B$-$\bar B$ mixing
and direct decay, for different modes mediated by 
$\,b\to s c \bar c$, $\,b\to s s \bar s$, and $b\to d c \bar c\,$ transitions. 
It is complementary in that it can simultaneously be regarded
as a test that ``polluting'' SM amplitudes are indeed as small
as expected, as well as a discriminant between different models
of new physics, should the deviations from the SM predictions
turn out to be large. 
These ideas find useful application in our discussion
of  $\,B\to K_S^{} K^+ K^-\,$  decay. 
One generally expects that physics beyond the standard model 
is most likely manifested through loop-level effects in 
$B$ decays~\cite{NP,GroWor}; new contributions may
well enter   
neutral-$B$ mixing or the penguin part of the decay amplitudes, or both.   
Consequently, in the 
CP-violating observables we consider, 
new physics is most likely relegated to the penguin terms. In this regard, 
the determination of  ${\cal A}_{D_s^{}\bar D_s^{}\pi}^{+-}$ in 
$B_s$ decay may be 
the most sensitive discriminant of emergent phenomena, as its asymmetry
is expected, on general grounds, to be very small in the SM.

\bigskip
   
\noindent{\bf Acknowledgments}\,\,
S.G. and J.T. are supported by
the U.S. Department of Energy under contracts 
DE-FG02-96ER40989 and DE-FG01-00ER45832.  
The work of J.T. was also supported by the Lightner-Sams Foundation.   
We thank J.A. Oller for the use of his scalar form factor program
and T. Gershon and H. Quinn for helpful comments. 
S.G. thanks the Institute for Nuclear Theory at the University
of Washington, Seattle and the SLAC Theory Group 
for hospitality during the completion
of this work.

\bigskip\bigskip
   
\appendix  

\noindent{\bf APPENDIX}  
 
\bigskip

For the  $\,B\to\rho\pi\,$  amplitudes in Eq.~(\ref{M_pirho}),  
the relevant matrix elements in the framework of Ref.~\cite{AliKL}
are  
\begin{eqnarray}   \label{0->pi}
\begin{array}{c}   \displaystyle      
\bigl\langle \pi^+(p) \bigl| \bar u\gamma_\mu^{} L d \bigr| 0 \bigr\rangle  
\,=\,  
\bigl\langle \pi^-(p) \bigl| \bar d\gamma_\mu^{} L u \bigr| 0 \bigr\rangle  
\,=\,  {\rm i} f_\pi^{}\, p_\mu^{}   \,\,,      
\vspace{1ex} \\   \displaystyle  
\bigl\langle \rho^+(p,\varepsilon) \bigr| 
\bar u\gamma_\mu^{} d\bigl| 0 \bigr\rangle   
\,=\,  
\bigl\langle \rho^-(p,\varepsilon) \bigr| 
\bar d\gamma_\mu^{} u \bigl| 0 \bigr\rangle   
\,=\,  f_\rho^{} M_\rho^{}\, \varepsilon_\mu^*   \,\,,    
\end{array}   
\end{eqnarray}   
\begin{eqnarray}  \label{B->r,p} 
\begin{array}{c}   \displaystyle      
q^\mu \bigl\langle \rho^+(p,\varepsilon) \bigr| 
\bar u \gamma_\mu^{} L b \bigl| \bar B^0(k) \bigr\rangle  
\,=\,  
-2{\rm i} A_0^{B\to\rho}(q^2) \, M_\rho^{}\, \varepsilon^\ast\cdot q    \,\,,  
\vspace{1ex} \\   \displaystyle  
\bigl\langle \pi^+(p) \bigr| \bar u \gamma^\mu L b
\bigl| \bar B^0 (k)\bigr\rangle  
\,=\,  
(k+p)^\mu\, F_1^{B\to\pi} (q^2) 
\,+\,  
{M_B^2-M_\pi^2\over q^2}\, q^\mu 
\left( f_0^{B\to\pi}(q^2)-F_1^{B\to\pi}(q^2) \right)   \,\,,  
\end{array}   
\end{eqnarray}   
where  $L \equiv 1 - \gamma_5^{}$, 
$q  \,\equiv\, k - p$, 
$f_\pi^{}$  and  $f_\rho^{}$  are the usual decay constants, 
and  $A_0^{}(q^2)$ and $F_{0,1}^{}(q^2)$  are form factors.  
In our phase convention for the meson flavor wave-functions, 
$\,\bar B^0=b\bar d,\,$  $\,\pi^+=u\bar d,\,$  
$\,\sqrt2\,\pi^0=u\bar u-d\bar d,\,$  $\,\pi^-=d\bar u,\,$, with 
the same convention for the  $\rho$.  
Then, using isospin symmetry, we also have  
$\,\bigl\langle \pi^+ \bigl| \bar u\gamma^\mu L d \bigr| 0 \bigr\rangle  
=\sqrt 2\, 
\bigl\langle \pi^0 \bigl| \bar u\gamma^\mu L u \bigr| 0 \bigr\rangle\,$  
and 
$\, \bigl\langle \pi^+ \bigr| \bar u\gamma^\mu b
\bigl| \bar B^0 \bigr\rangle 
=-\sqrt 2\, \bigl\langle \pi^0 \bigr| 
\bar d\gamma^\mu b \bigl| \bar B^0 \bigr\rangle .\,$  
The resulting amplitudes for  $\,\bar B^0\to\rho\pi\,$  are  
\begin{eqnarray}    
\begin{array}{c}   \displaystyle   
\bar a_{+-}^{}  \,=\,     
\sqrt 2\, G_{\rm F}^{} \left\{   
\lambda_u^{(d)}\, a_1^{}   
- \lambda_t^{(d)} \left[ \vphantom{|_|^|}
a_4^{}+a_{10}^{} - \bigl( a_6^{}+a_8^{} \bigr) \, R_{\pi^-}^{} \right]   
\right\} f_\pi^{} M_\rho^{}\, A_0^{B\to\rho}\bigl(M_\pi^2\bigr)   \,\,,  
\vspace{2ex} \\   \displaystyle   
\bar a_{-+}^{}  \,=\,  
\sqrt 2\, G_{\rm F}^{} \left[  
\lambda_u^{(d)}\, a_1^{} 
- \lambda_t^{(d)}\, \bigl( a_4^{}+a_{10}^{} \bigr)    
\right] f_\rho^{} M_\rho^{}\, F_1^{B\to\pi}\bigl(M_\rho^2\bigr)   \,\,,   
\vspace{2ex} \\   \displaystyle   
\bar a_{00}^{}  \,=\,    
{-G_{\rm F}^{}\over \sqrt 2} \left\{ 
\lambda_u^{(d)}\, a_2^{} 
+ \lambda_t^{(d)} \left[ \vphantom{|_|^|}
a_4^{} + \mbox{$\frac{3}{2}$} a_7^{} - \mbox{$\frac{3}{2}$} a_9^{} 
- \mbox{$\frac{1}{2}$} a_{10}^{} 
- \left( a_6^{}-\mbox{$\frac{1}{2}$} a_8^{} \right) R_{\pi^0}^{} 
\right] \right\} f_\pi^{} M_\rho^{}\, A_0^{B\to\rho}\bigl(M_\pi^2\bigr)  
\vspace{1ex} \\   \displaystyle   
-\,\,    
{G_{\rm F}^{}\over \sqrt 2} \left[   
\lambda_u^{(d)}\, a_2^{} 
+ \lambda_t^{(d)} \left( \vphantom{|_|^|}
a_4^{} - \mbox{$\frac{3}{2}$} a_7^{} - \mbox{$\frac{3}{2}$} a_9^{} 
- \mbox{$\frac{1}{2}$} a_{10}^{} 
\right) \right]  
f_\rho^{} M_\rho^{}\, F_1^{B\to\pi}\bigl(M_\rho^2\bigr)   \,\,,   
\hspace*{4em} 
\end{array}     
\end{eqnarray}     
where the numerical values of  $a_i^{}$ for 
$\,b\to d\,$ decay  are obtained from 
Table~\ref{a_i}, 
$\,R_{\pi^-}^{}\equiv 2M_{\pi^-}^2/\bigl[\bigl(m_u^{}+m_b^{}\bigr)
\bigl(m_u^{}+m_d^{}\bigr)\bigr],\,$ 
and  
$\,R_{\pi^0}^{}\equiv M_{\pi^0}^2/\bigl[\bigl(m_d^{}+m_b^{}\bigr)
m_d^{}\bigr].\,$    
The amplitudes $a_{kl}^{}$  for $\,B^0\to\rho\pi\,$  can be 
found from  $\bar a_{lk}^{}$  by replacing  
$\lambda_{q'}^{(q)}$  with  $\lambda_{q'}^{(q)*}$   and 
using the  $\bar b\to\bar d$  entries in Table~\ref{a_i}.

Turning to the $\,B\to\rho\pi\to3\pi\,$  amplitudes in 
Eqs.~(\ref{M_3pi},\ref{f_cdn}),\footnote{
The signs of the different 
terms in Eq.~(\ref{f_cdn}) 
follow from  the $\rho\to\pi\pi$  
couplings  
$\,\bigl\langle\pi^0\bigl(p_0^{}\bigr)\,   
\pi^\pm\bigl(p_\pm^{}\bigr)\big| \rho^\pm\bigr\rangle  
=\pm g_{\rho\pi\pi}^{}\, 
\varepsilon_\rho^{}\cdot \bigl( p_0^{}-p_\pm^{} \bigr) \,$  
and  
$\,\bigl\langle\pi^+\bigl(p_+^{}\bigr)\,   
\pi^-\bigl(p_-^{}\bigr) \big| \rho^0\bigr\rangle 
=g_{\rho\pi\pi}^{}\, 
\varepsilon_\rho^{}\cdot \bigl( p_+^{}-p_-^{} \bigr) ,\,$  
where we adopt the notation
$\,\bigl\langle M_2^{} M_3^{}\bigr|M_1^{}\bigr\rangle\equiv
\bigl\langle M_2^{} M_3^{}\bigr|{\cal H}_{\rm strong}^{}
\bigr|M_1^{}\bigr\rangle.\,$  
The signs of these couplings follow, in turn, from 
the phase convention we have chosen for the flavor wave-functions: 
$\,\bigl|\pi^\pm \bigr\rangle=\mp \bigl|I=1,I_3=\pm 1\bigr\rangle\,$
and  $\,\bigl|\pi^0 \bigr\rangle = \bigl|I=1,I_3=0 \bigr\rangle,\,$ etc.
}  
we note that they contain 
the $\rho\to\pi\pi$ vertex function  $\Gamma_{\rho\pi\pi}^{}(s)$,  
which is given by
\begin{eqnarray}   \label{Grpp}
\Gamma_{\rho\pi\pi}^{}(s)  \,=\,   
{-F_\rho^{}(s)\over f_{\rho\gamma}^{}}   \,\,,  
\end{eqnarray}   
where $F_\rho^{}(s)$  is the vector form-factor of the pion 
and  $f_{\rho\gamma}^{}$  is the  $\rho$-$\gamma$ coupling  
constant.   
The form factor $F_\rho^{}(s)$ is determined by fitting to  
$\,e^+e^-\to\pi^+\pi^-\,$  data with a parametrization consistent with 
theoretical constraints~\cite{GarOCon1}.
The value of $f_{\rho\gamma}^{}$ is determined from the 
$\,\rho\to e^+e^-\,$  width, which, in turn, is extracted from 
the  $\,e^+e^- \to \pi^+\pi^-\,$  cross section at  
$\,s=M_\rho^2\,$~\cite{GarOCon2}.    
The overall sign is chosen so that Eq.~(\ref{Grpp}) is consistent with 
the Breit-Wigner (BW) form, 
\begin{eqnarray}   \label{GrppBW}
\Gamma_{\rho\pi\pi}^{\rm BW}(s)  \,=\,   
{g_{\rho\pi\pi}^{}\over s-M_\rho^2+{\rm i}\Gamma_\rho^{} M_\rho^{}}  \,\,  
\end{eqnarray}   
as $\,s\to M_\rho^2,\,$  where  $g_{\rho\pi\pi}^{}$  is 
the  $\,\rho\to\pi\pi\,$  coupling constant.    
In our numerical analysis, 
we employ the ``solution $B$'' fit of  Ref.~\cite{GarOCon1} for 
$F_\rho^{}(s)$, for which  
$\,f_{\rho\gamma}^{}=0.122 \pm 0.001\,\rm GeV^2\,$~\cite{GarOCon2}.
For $s\ge (M_\pi+  M_\omega)^2$ 
we match  $\Gamma_{\rho\pi\pi}^{}(s)$  to the BW expression, as
the fit for $F_\rho(s)$ exists exclusively in the region where
$\pi\pi$ scattering is elastic, see Refs.~\cite{GarMei,gt} for
further details. 

\begin{table}[tb]   
\caption{\label{a_i}%
Numerical values of the coefficients  $a_i^{}$ with 
$\,N_{\rm c}^{}=3,\,$  reproduced from Tables~VI and~VII in 
Ref.~\protect{\cite{AliKL}},  for
$\,b\to q$ and $\bar b\to\bar q$  transitions, with  $\,q=d,s.\,$    
Each of the entries for  $\,a_3^{},\cdots,a_{10}^{}\,$   
ought to be multiplied by a factor of $10^{-4}$.   
} \centering   \footnotesize
\vskip 0.5\baselineskip
\begin{tabular}{@{\hspace{3ex}}l@{\hspace{5ex}}c@{\hspace{5ex}}
c@{\hspace{5ex}}c@{\hspace{5ex}}c@{\hspace{3ex}}}  
\hline \hline  
  &  $b\to d\vphantom{\Bigl|}$  &  $\bar b\to\bar d$  &  
$b\to s$  &  $\bar b\to\bar s$  
\\ \hline && && \vspace{-3ex} \\   
$a_1^{}$  &  1.05  &  1.05  &  1.05  &  1.05  \\  
$a_2^{}$  &  0.053  &  0.053  &  0.053  &  0.053  \\  
$a_3^{}$  &  48    &  48    &  48    &  48    \\  
$a_4^{}$  &  $-412-36\,\rm i$  &  $-461-124\,\rm i$  &  $-439-77\,\rm i$  &  
$-431-77\,\rm i$  \\   
$a_5^{}$  &  $-45$  &  $-45$  &  $-45$  &  $-45$  \\  
$a_6^{}$  &  $-548-36\,\rm i$  &  $-597-124\,\rm i$  &  $-575-77\,\rm i$  &  
$-568-77\,\rm i$  \\   
$a_7^{}$  &  $0.7-1.0\,\rm i$  &  $0.3-1.8\,\rm i$  &  $0.5-1.3\,\rm i$  &  
$0.5-1.3\,\rm i$ \\  
$a_8^{}$  &  $4.7-0.3\,\rm i$  &  $4.5-0.6\,\rm i$  &  $4.6-0.4\,\rm i$  &  
$4.6-0.4\,\rm i$  \\  
$a_9^{}$  &  $-94-1.0\,\rm i$  &  $-95-1.8\,\rm i$  &  $-94-1.3\,\rm i$  &   
$-94-1.3\,\rm i$ \\  
$a_{10}^{}$  &  $-14-0.3\,\rm i$  &  $-14-0.6\,\rm i$  &  $-14-0.4\,\rm i$  &  
$-14-0.4\,\rm i$  \\  
&& && \vspace{-3ex} \\   
\hline \hline  
\end{tabular}
\vspace{1ex} \\  
\end{table}

To incorporate the  $\,B\to\sigma\pi^0\,$  contributions,  
we need the additional matrix element 
\begin{eqnarray}  \label{B->s} 
\begin{array}{c}   \displaystyle      
q^\mu \bigl\langle \sigma(p) \bigr| 
\bar d\gamma_\mu^{}L b \bigl| \bar B^0(k) \bigr\rangle  
\,=\,   
-{\rm i} \left( M_B^2-M_\sigma^2 \right) f_0^{B\to\sigma}(q^2)  \,\,.  
\end{array}   
\end{eqnarray}   
It follows that the  $\,\bar B^0\to\sigma\pi^0\,$  amplitude 
in Eq.~(\ref{M_pisigma}) is 
\begin{eqnarray}    
\bar a_\sigma^{}  &=&    
{G_{\rm F}^{}\over 2} \left\{   
\lambda_u^{(d)}\, a_2^{}   
+ \lambda_t^{(d)} \left[ \vphantom{|_|^|}  
a_4^{} + \mbox{$\frac{3}{2}$} a_7^{} 
- \mbox{$\frac{3}{2}$} a_9^{} - \mbox{$\frac{1}{2}$} a_{10}^{} 
- \left( a_6^{}-\mbox{$\frac{1}{2}$} a_8^{} \right) R_{\pi^0}^{} 
\right] \right\} \left( M_{B^0}^2-M_\sigma^2 \right) 
f_\pi^{} F_0^{B\to\sigma}\bigl(M_\pi^2\bigr) 
\nonumber \\ && \!\!\!  
-\,\,   
G_{\rm F}^{}\, \lambda_t^{(d)}\, 
\left( a_6^{}-\mbox{$\frac{1}{2}$} a_8^{} \right)   
{\langle\sigma|\bar d d|0\rangle\over m_b^{}-m_d^{}}   
\left( M_{B^0}^2-M_{\pi^0}^2 \right) 
F_0^{B\to\pi} \bigl(M_\sigma^2\bigr)   \,\,, 
\end{eqnarray}     
where we use  $\,M_\sigma^{}=478\,{\rm MeV}\,$ determined in 
Ref.~\cite{e791},  
$\,F_0^{B\to\sigma}\bigl(M_\pi^2\bigr)=0.46\,$ as per 
Refs.~\cite{DeaPol}, 
and
$\,\langle\sigma|\bar d d|0\rangle= 
2 M_{\pi^+}^2/\bigl[\sqrt 6\,\chi\, \bigl(m_u^{}+m_d^{}\bigr)\bigr],\,$   
with  $\,\chi=20.0\,{\rm GeV}^{-1}\,$~\cite{GarMei}.     
To obtain the  $\,\bar B^0\to\sigma\pi^0\,$  amplitude, we note that   
$\,CP\bigl|\rho^0\pi^0\bigr\rangle=+|\rho^0\pi^0\bigr\rangle\,$  
and  
$\,CP\bigl|\sigma\pi^0\bigr\rangle=-|\sigma\pi^0\bigr\rangle\,$  
for  $\,B\to\rho\pi,\sigma\pi,\,$  as the relative angular-momenta 
in the final states are  $\,l=1$ and $0,\,$  respectively.  
Consequently,  $a_\sigma^{}$  can be found from  
$\bar a_\sigma^{}$ by replacing  $\lambda_{q'}^{(q)}$  with  
$\lambda_{q'}^{(q)*}$,  using the  $\bar b\to\bar d$  entries 
in Table~\ref{a_i},  and adding an overall minus sign,  
which is consistent with  Eq.~(\ref{CP[a_s]}).

The  $\sigma$  contribution to the  $\,B\to 3\pi\,$  
amplitudes in Eq.~(\ref{M_3pi'}) involve the function  
$\,f_\sigma^{}(s)=\Gamma_{\sigma\pi\pi}^{}(s),\,$  which 
describes the $\sigma \to \pi\pi$  vertex function in the isospin  $\,I=0\,$  
and  angular-momentum  $\,J=0\,$  channel.   
This function must satisfy known theoretical 
constraints~\cite{GarMei}, and we adopt 
\begin{eqnarray}   \label{Gspp}  
\Gamma_{\sigma\pi\pi}^{}(s)  \,=\,  \chi\, \Gamma_1^{n*}(s)   \,\,,  
\end{eqnarray}   
for our numerical work, 
where the function  $\Gamma_1^n(s)$  is computed in 
Ref.~\cite{MeiOll}.  
We remark that the theoretically consistent  
$\Gamma_{\sigma\pi\pi}^{}(s)$  is very different from the 
Breit-Wigner form used in Ref.~\cite{e791}, as detailed in 
Ref.~\cite{GarMei}.

For the  $\,B\to D^{*\pm} D^\mp\,$  amplitudes in  Eq.~(\ref{M_D*D}),  
the needed matrix elements are analogous to those  
of the  $\rho\pi$  case, where the replacement of the $u$ with $c$
quarks is realized throughout. 
Thus for  $\,\bar B^0\to D^*D\,$  decay, we have 
\begin{eqnarray}    
\begin{array}{c}   \displaystyle   
\bar a_{+-}^{}  \,=\,     
\sqrt 2\, G_{\rm F}^{} \left\{   
\lambda_c^{(d)}\, a_1^{}   
- \lambda_t^{(d)} \left[ \vphantom{|_|^|}
a_4^{}+a_{10}^{} - \bigl( a_6^{}+a_8^{} \bigr) \, R_D^{} \right]   
\right\} f_D^{} M_{D^*}^{}\, A_0^{B\to D^*}\bigl(M_D^2\bigr)   \,\,,  
\vspace{2ex} \\   \displaystyle   
\bar a_{-+}^{}  \,=\,  
\sqrt 2\, G_{\rm F}^{} \left[  
\lambda_c^{(d)}\, a_1^{} 
- \lambda_t^{(d)}\, \bigl( a_4^{}+a_{10}^{} \bigr)    
\right] f_{D^*}^{} M_{D^*}^{}\, F_1^{B\to D}\bigl(M_{D^*}^2\bigr)   \,\,,   
\end{array}     
\end{eqnarray}     
where 
$\,R_D^{}\equiv 2M_D^2/\bigl[\bigl(m_c^{}+m_b^{}\bigr)
\bigl(m_c^{}+m_d^{}\bigr)\bigr].\,$ 
For the form factors,  we use~\cite{iwf} 
\begin{eqnarray}   
2\, F_1^{B\to D} \bigl( q^2 \bigr)  \,=\,   
{m_B^{}+m_D^{}\over \sqrt{m_B^{}m_D^{}}}\, \xi(w)   \,\,,  
\hspace{3em} 
2\, A_0^{B\to D^*} \bigl( q^2 \bigr)  \,=\,   
{m_B^{}+m_{D^*}^{}\over \sqrt{m_B^{}m_{D^*}^{}}}\, \xi(w)   \,\,,   
\end{eqnarray}     
where
\begin{eqnarray}   
\xi(w)  \,=\,  \left( {2\over 1+w} \right) ^2   \,\,,   
\hspace{3em}  
w  \,=\,  {m_B^2+m_{D^{(*)}}^2-q^2\over 2\, m_B^{}\, m_{D^{(*)}}^{}}   \;.  
\end{eqnarray}     
The amplitudes $a_{kl}^{}$  for $\,B^0\to D^*D\,$  decay are obtained 
from  $\bar a_{lk}^{}$ by replacing  
$\lambda_{q'}^{(q)}$  with  $\lambda_{q'}^{(q)*}$   and 
using the  $\bar b\to\bar d$  entries in Table~\ref{a_i}.

The  $\,B\to D^\pm D^\mp\pi^0\,$  amplitudes in  Eq.~(\ref{M_DDpi})  
contain the functions  
\begin{eqnarray}   \label{f+-}
f_\pm^{}  \,=\,   
{ \pm\mbox{$\frac{1}{2}$}\, g_{D^*D\pi}^{} 
 \over  s_{\pm0}^{}-m_{D^*}^2+{\rm i}\Gamma_{D^*}^{} m_{D^*}^{} }   
\left[ s_{+-}^{}-s_{\mp0}^{}-m_D^2+m_\pi^2 
      - \left( m_{B^{}}^2-s_{\pm0}^{}-m_D^2 \right) 
       {m_D^2-m_\pi^2\over m_{D^*}^2} \vphantom{\int_|^|} \right]   \,\,,    
\end{eqnarray}     
where $g_{D^*D\pi}^{}$  is the coupling constant 
associated with 
the strong decays 
$\,D^{*\pm}\to D^{\pm}\pi^0,\,$  
\begin{eqnarray}   \label{M[D^*Dp])}
{\cal M}_{D^{*\pm}\to D^\pm\pi^0}^{}  \,=\,  
\pm 2\, g_{D^*D\pi}^{}\, \varepsilon\!\cdot\! p_\pi^{}   \;.     
\end{eqnarray}     
From the empirical values 
$\,{\cal B} \bigl( D^{*+}\to D^+\pi^0 \bigr)=(30.7\pm0.5)\%\,$~\cite{pdb}  
and
$\,\Gamma_{D^{*+}}^{}=(96\pm4\pm22)\rm keV,\,$~\cite{D*width}  
we extract  $\,g_{D^*D\pi}^{}=6.3.\,$   
The use of the Breit-Wigner form for the $D^*$ 
resonance in Eq.~(\ref{f+-}) is appropriate as the $D^*$ is
narrow.

The  $\,B_s^{}\to K^{*\pm} K^\mp\,$  amplitudes also follow from
the  $\rho\pi$  case,  and  thus we have for  
$\,\bar B_s^0\to K^{*\pm}K^\mp\,$   
\begin{eqnarray}    
\begin{array}{c}   \displaystyle   
\bar a_{+-}^{}  \,=\,     
\sqrt 2\, G_{\rm F}^{} \left\{   
\lambda_u^{(s)}\, a_1^{}   
- \lambda_t^{(s)} \left[ \vphantom{|_|^|}
a_4^{}+a_{10}^{} - \bigl( a_6^{}+a_8^{} \bigr) \, R_K^{} \right]   
\right\} f_K^{} M_{K^*}^{}\, A_0^{B\to K^*}\bigl(M_K^2\bigr)   \,\,,  
\vspace{2ex} \\   \displaystyle   
\bar a_{-+}^{}  \,=\,  
\sqrt 2\, G_{\rm F}^{} \left[  
\lambda_u^{(s)}\, a_1^{} 
- \lambda_t^{(s)}\, \bigl( a_4^{}+a_{10}^{} \bigr)    
\right] f_{K^*}^{} M_{K^*}^{}\, F_1^{B\to K}\bigl(M_{K^*}^2\bigr)   \,\,,   
\end{array}     
\end{eqnarray}     
where the  $a_i^{}$  values for $\,b\to s\,$ decay are given 
in Table~\ref{a_i}  and    
$\,R_K^{}\equiv 2M_K^2/\bigl[\bigl(m_u^{}+m_b^{}\bigr)
\bigl(m_u^{}+m_s^{}\bigr)\bigr].\,$ 
The amplitudes $a_{kl}^{}$  for $\,B_s^0\to K^{*\pm}K^\mp\,$ follow 
from  $\bar a_{lk}^{}$  by replacing  
$\lambda_{q'}^{(q)}$  with  $\lambda_{q'}^{(q)*}$   and 
using the  $\bar b\to\bar s$  entries in Table~\ref{a_i}.  
The  $\,K^{*\pm}\to K^\pm\pi^0\,$  coupling constant occurring 
in the analogue of  $f_\pm^{}$ in Eq.~(\ref{f+-}) is 
$\,g_{K^*K\pi}^{}=3.2,\,$  which is extracted using the measured  $K^{*+}$  
width~\cite{pdb}  and  the isospin relation  
$\,\Gamma_{K^{*+}\to K\pi}^{}=3\,\Gamma_{K^{*+}\to K^+\pi^0}^{}.\,$

Finally, we list all the input parameters which have not been 
previously mentioned. 
For the heavy mesons, we have
\begin{eqnarray}   
\begin{array}{c}   \displaystyle   
M_{B^0}^{}  \,=\,  5.2794\,{\rm GeV}   \,\,,   
\hspace{2em}     
\tau_{B^0}^{}  \,=\,  1.548\times 10^{-12}\,{\rm s}  \,\,,   
\vspace{1ex} \\   \displaystyle   
M_{B_s^0}^{}  \,=\,  5.3696\,{\rm GeV}   \,\,,   
\hspace{2em}     
\tau_{B_s^0}^{}  \,=\,  1.493\times 10^{-12}\,{\rm s}  \,\,,   
\vspace{1ex} \\   \displaystyle   
f_D^{}  \,=\,  200\,{\rm MeV}   \,\,,   \hspace{1em} 
M_{D^\pm}^{} \,=\,  1869.3\,{\rm MeV}   \,\,,   
\hspace{2em}  
f_{D^*}^{}  \,=\,  230  \,{\rm MeV}  \,\,,   \hspace{1em}  
M_{D^{*\pm}}^{}  \,=\,  2010.0\,{\rm MeV}  \,\,,  
\vspace{1ex} \\   \displaystyle   
f_{D_s^{}}^{}  \,=\,  240\,{\rm MeV}   \,\,,   \hspace{1em} 
M_{D_s^\pm}^{} \,=\,  1968.6\,{\rm MeV}   \,\,,   
\hspace{2em}     
f_{D_s^*}^{}  \,=\,  275  \,{\rm MeV}  \,\,,   \hspace{1em}  
M_{D_s^{*\pm}}^{}  \,=\,  2112.4\,{\rm MeV}  \,\,,  
\end{array}      
\end{eqnarray}      
and for the light mesons 
\begin{eqnarray}   
\begin{array}{c}   \displaystyle   
f_\pi^{}  \,=\,  134\,{\rm MeV}   \,\,,   \hspace{2em} 
M_{\pi^\pm}^{} \,=\,  139.57\,{\rm MeV}   \,\,,   \hspace{2em} 
M_{\pi^0}^{} \,=\,  134.98\,{\rm MeV}   \,\,,  
\vspace{1ex} \\   \displaystyle   
f_\rho^{}  \,=\,  210  \,{\rm MeV}  \,\,,   \hspace{2em}  
M_\rho^{}  \,=\,  769.3\,{\rm MeV}  \,\,,  \hspace{2em}     
\Gamma_\rho^{}  \,=\,  150\,{\rm MeV}   \,\,,  \hspace{2em}  
g_{\rho\pi\pi}^{}  \,=\,  5.8   \,\,,  
\vspace{1ex} \\   \displaystyle   
f_K^{}  \,=\,  158\,{\rm MeV}   \,\,,   \hspace{2em} 
M_{K^\pm}^{} \,=\,  493.68\,{\rm MeV}   \,\,,   \hspace{2em} 
\vspace{1ex} \\   \displaystyle   
f_{K^*}^{}  \,=\,  214\,{\rm MeV}  \,\,,   \hspace{2em}  
M_{K^{*\pm}}  \,=\,  891.66\,{\rm MeV}  \,\,,  \hspace{2em}     
\Gamma_{K^{*\pm}}^{}  \,=\,  50.8\,{\rm MeV}   \,\,,  
\end{array}      
\end{eqnarray}      
where the decay constants have been obtained from Ref.~\cite{NeuSte}. 
For current quark masses, we use  
the running values at  $\,\mu=2.5\,\rm GeV\,$  found in 
Ref.~\cite{AliKL}:
\begin{eqnarray}   
\begin{array}{c}   \displaystyle   
m_u^{}  \,=\,  4.2\,{\rm MeV}   \,\,,   \hspace{2em}     
m_d^{}  \,=\,  7.6\,{\rm MeV}   \,\,,   \hspace{2em}     
m_s^{}  \,=\,  122\,{\rm MeV}   \,\,,   
\vspace{1ex} \\   \displaystyle   
m_c^{}  \,=\,  1.5\,{\rm GeV}   \,\,,   \hspace{2em}     
m_b^{}  \,=\,  4.88\,{\rm GeV}   \,\,,   
\end{array}      
\end{eqnarray}      
For the heavy-to-light form factors, we employ 
\begin{eqnarray}   
\begin{array}{c}   \displaystyle   
A_0^{B\to\rho}\bigl(M_\pi^2\bigr)  \,\simeq\,  
A_0^{B\to\rho}(0)  \,=\,  0.28   \,\,,  
\hspace{2em}  
F_1^{B\to\pi}\bigl(M_\rho^2\bigr)  \,\simeq\,
F_1^{B\to\pi}(0)  \,=\,  0.33   \,\,,  
\vspace{1ex} \\   \displaystyle   
F_0^{B\to\pi}\bigl(M_\sigma^2\bigr)  \,\simeq\,
F_0^{B\to\pi}(0)  \,=\,  0.33   \,\,,  
\vspace{1ex} \\   \displaystyle   
A_0^{B_s^{}\to K^*}\bigl(M_K^2\bigr)  \,\simeq\,  
A_0^{B_s^{}\to K^*}(0)  \,=\,  0.24   \,\,,  
\hspace{2em}  
F_1^{B_s^{}\to K}\bigl(M_{K^*}^2\bigr)  \,\simeq\,
F_1^{B_s^{}\to K}(0)  \,=\,  0.27   \,\,,  
\end{array}     
\end{eqnarray}     
estimated in the  BSW  model~\cite{bsw}.


\end{document}